\documentclass[iop, revtex4]{emulateapj}
\usepackage{graphicx}
\usepackage{amstext}
\usepackage{amsmath}
\usepackage{xspace}
\usepackage{color}
\usepackage{rotating}
\usepackage{txfonts}
\usepackage{booktabs}
\usepackage{url}
\usepackage{threeparttable}
\usepackage{longtable}
\usepackage{lscape}
\bibpunct{(}{)}{;}{a}{}{,}

\newcommand{\Msun}{\ensuremath{M_\odot}\xspace}
 
\newcommand{\Teffsun}{\ensuremath{{T_{\text{eff}\,\odot}}}\xspace}
\newcommand{\loggsun}{\ensuremath{{\log g}_\odot}\xspace}
\newcommand{\Teff}{\ensuremath{{T_{\text{eff}}}}\xspace}
\newcommand{\logg}{\ensuremath{{\log g}}\xspace}
\newcommand{\loggf}{\ensuremath{{\log gf}}\xspace}
\newcommand{\vt}{\ensuremath{{v_t}}\xspace}
\graphicspath{
{./plots/}
}

\begin{document}
\title{Carbon and Oxygen abundances across the Hertzsprung gap}

\author{Jens Adamczak and David L. Lambert}
\email{adamczak@astro.as.utexas.edu, dll@astro.as.utexas.edu}
\affil{McDonald Observatory, The University of Texas, Austin, Texas, 78712, USA}
 
\begin{abstract}
We derived atmospheric parameters and spectroscopic abundances for C and O for a large sample of
stars located in the Hertzsprung gap in the Hertzsprung-Russell Diagram in order to detect chemical
peculiarities and get a comprehensive overview of the population of stars in this evolutionary
state.  We have observed and analyzed high resolution spectra (R = 60\,000) of 188 stars in the mass
range 2 -- 5\,\Msun with the 2.7\,m Harlan J. Smith Telescope at the McDonald Observatory including
28 stars previously identified as Am/Ap stars.  We find that the C and O abundances of the majority
of stars in the Hertzsprung gap are in accordance with abundances derived for local lower mass
dwarfs but detect expected peculiarities for the Am/Ap stars. The C and O abundances of stars
with $\Teff < 6500\,$K are slightly lower than for the hotter objects but the C/O ratio is constant in
the analyzed temperature domain. No indication of an alteration of the C and O abundances of the
stars by mixing during the evolution across the Hertzsprung gap could be found before the
homogenization of their atmospheres by the first dredge-up. 

\end{abstract}

\section{Introduction}
The chemical evolution of normal mass stars from the main sequence (MS) to the
ascent of the red giant branch (RGB) in the Hertzsprung-Russell Diagram (HRD)
is complex and involves a sequence of processes that alter the composition and
structure of their atmospheres. A key facet is the development of an expanding
convective envelope that connects the outer layers of the atmosphere with
deeper regions of nuclear burning. In the process the atmospheric abundances of
several elements are altered and result in the abundance patterns typical for
giant stars, i.e. a slight C deficiency that is accompanied by a N
overabundance and a strong depletion of Li. Observational investigations
consistently confirm these abundance patterns and the altered abundances of
giants after the first dredge-up can be qualitatively predicted from the
abundances of the corresponding normal mass dwarf progenitors.

Main sequence stars in the mass range 2 -- 5\,\Msun carry characteristics not exhibited by stars of
about 1\,\Msun which provide the majority of G and K giants. The higher mass stars have a larger
range in rotational velocities whereas solar mass main sequence stars are all slow rotators; a sharp
break in projected rotational velocity occurs at about 1.3\,\Msun
\nocite{1967ApJ...150..551K}({Kraft} 1967). In addition, the main sequence domain of the higher mass
stars is populated with a variety of chemically-peculiar stars of spectral types Ap and Am for which
diffusion of elements is considered to have distorted the composition of their atmospheres. A
different group of chemically-peculiar stars can be found on the other side of the Hertzsprung gap
in the form of weak G band stars (wk Gb stars), a rare group of early G and K giants of slightly
sub-solar metallicity that show unusually large underabundances of C. While the origin of the
chemical peculiarities of the Am/Ap stars can be explained reasonably well by theoretical models,
the exact cause for the wk Gb abundance anomalies is still subject to speculation. 

For these stars in the higher mass range, the crossing of the Hertzsprung gap marks an essential
period in their evolution and the mechanisms that affect the distribution of elements in their
atmospheres, culminating in the first dredge-up, effectively determine the abundance patterns of
their descendants on the giant branch.  However, apart from determinations of abundances for Li and
selected elements \nocite{1994AJ....107.2211W, 1996PhDT........75H,
2003A&A...410..937D}({Wallerstein} {et~al.} 1994; {Hiltgen} 1996; {de Laverny} {et~al.} 2003), to
our knowledge, no extensive abundance analysis of a larger number of stars in this evolutionary
stage has been performed yet. Especially the abundances of the basic elements C, N, and O are of
interest, since these elements are altered by the nuclear processes involved in the energy
production via the CNO-cycle and are different for dwarf and giant stars. These elements are
therefore the key indicators to analyse abundance patterns, identify and investigate chemical
peculiarities, and get a comprehensive picture of the characteristics of stars in the Hertzsprung
gap.  The determination of abundances of these and other elements necessarily faces the difficulty
that spectra of many stars will be greatly rotationally broadened thus complicating, even thwarting,
an abundance analysis.  Nonetheless, in this paper, we aim to provide a detailed analysis of C and O
abundances\footnote{We do not consider N due to the lack of suitable lines in the recorded spectral
range.} of a large sample of stars in the mass range 2 -- 5\,\Msun that are located in the
Hertzsprung gap in the HRD. 

\section{Observations and target selection}\label{sect:observations} 
We compiled a sample of target stars from the Hipparcos catalog \nocite{2007A&A...474..653V}({van
Leeuwen} 2007). The primary selection criterion was their position in the Hertzsprung-Russell
Diagram (HRD). Effective temperatures were calculated from B-V colors with the calibrations from
\nocite{1999A&AS..140..261A}{Alonso}, {Arribas}, \&  {Mart{\'{\i}}nez-Roger} (1999) and luminosities
were determined with the Hipparcos parallaxes and V magnitudes. We selected stars with masses
between 2 and 5\,\Msun by comparison with evolutionary tracks from \nocite{2008A&A...484..815B,
2009A&A...508..355B}{Bertelli} {et~al.} (2008, 2009) and a lower temperature limit of $\log (\Teff)
> \log (3.72)$. The sample was
furthermore restricted to stars with small errors in their parallaxes and declinations $>
-20^\circ$. For a first set of observations, stars with an error in their parallaxes of less or equal
than 20\% of the actual parallax and a right ascension (RA) in the range of 4 -- 20\,h were
observed. This sample was complemented with additional observations of objects with a parallax error
of less or equal than 50\% of the parallax in order to better cover the less populated area of
higher masses (see Section  \ref{sect:masses}). For the second sample, stars with a right ascension
between 20 -- 11\,h were accessible. It should be noted that with this target selection it is
impossible to detect the direction of the transition of a star through the Hertzsprung gap and to
distinguish, e.g. a main sequence star on its way to the giant branch from a pulsating horizontal
branch star or a post-AGB star.
 
We observed 188 stars with the 2.7\,m Harlan J. Smith Telescope at the McDonald Observatory.  The
telescope was equipped with the Robert G. Tull Cross-Dispersed Echelle Spectrograph
\nocite{1995PASP..107..251T}({Tull} {et~al.} 1995) and a Tektronix 2048\,x\,2048 pixel CCD detector.
The stars were observed in what we deem the standard (std) setup centered on 5060\,\AA\xspace in
order 69. This setup up was chosen in order to cover as many of the interesting spectral lines as
possible. The observations covered a wavelength range of 3\,900 -- 10\,000\,\AA\xspace with a
resolving power of $\text{R}\equiv \lambda/ \Delta \lambda = 60\,000$. The observations are listed
in Table \ref{tab:observations}. Unless noted otherwise the spectra were reduced in a multiple step
procedure. First flat field and bias corrections were applied. Then the echelle spectra were
extracted and wavelength calibrated using ThAr-lamp comparison exposures taken before or after the
observations. The different spectral orders were combined and continuum normalized. All tasks were
performed using the IRAF package \footnote{IRAF is distributed by the National Optical Astronomy
Observatories, which are operated by the Association of Universities for Research in Astronomy,
Inc., under cooperative agreement with the National Science Foundation.}. 
\tabletypesize{\small}
\begin{deluxetable}{llllrr}
\tablecaption{Observation log of analyzed spectra \label{tab:observations}.} \tablewidth{0pt}
\tablehead{
\colhead{HIP} & \colhead{HD} & \colhead{V [mag]} &\colhead{Obs. date} & \colhead{$t_{exp}$ [s]} & \colhead{SNR}
}
\startdata
HIP\,8066 & HD\,10497  & 6.9 & 2013 Nov 30 & 840         & 252 \\ 
HIP\,11115 & HD\,14542 & 7.1 & 2013 Nov 30 & 1020        & 232 \\ 
HIP\,13004 & HD\,17086 & 6.6 & 2013 Nov 30 & 660         & 190 \\ 
HIP\,13036 & HD\,17245 & 6.6 & 2013 Nov 30 & 720         & 284 \\ 
HIP\,16001 & HD\,21085 & 7.2 & 2013 Nov 30 & 1140        & 250 \\ 
HIP\,16203 & HD\,21483 & 7.1 & 2013 Nov 30 & 1080        & 175 
\enddata
\tablecomments{Table \ref{tab:observations} is published in its entirety in the electronic edition
of \textsl{The Astrophysical Journal}. A portion is shown here for guidance regarding its form and
content.}
\end{deluxetable}

\section{Parameter determination \label{sect:parameters}}
In order to determine reliable effective temperatures and atmospheric parameters we utilized a
variety of different techniques that are described in the following.

\subsection{Fe line equivalent widths}
Atmospheric parameters were derived from Fe line equivalent widths (EWs). We will refer to these
parameters as spectroscopic keeping in mind that some of the other methods also rely on the observed
spectra. The equivalent widths of 37 Fe\,{\sc i} and 9 Fe\,{\sc ii} lines were measured with the
IRAF routine splot whenever possible. Many of the stars are fast rotators. In some cases a
de-blending of the lines was necessary, in other cases no determination of equivalent widths was
possible at all. The \loggf values and excitation potentials were taken from
\nocite{2011ApJ...743..135R}{Ram{\'{\i}}rez} \& {Allende Prieto} (2011). The line list and measured
equivalent widths were taken as an input for the spectral synthesis code MOOG
\nocite{1973PhDT.......180S}({Sneden} 1973). One-dimensional (1D), plane-parallel input model
atmospheres in local thermodynamic equilibrium (LTE) were interpolated from a grid of models from
\nocite{2004astro.ph..5087C}{Castelli} \& {Kurucz} (2004). The abundances of the computed Fe lines
for a given set of parameters were force-fitted to match the measured EWs. NLTE effects that could
alter the determination of Fe abundances and thus affect the stellar parameters are negligible for
stars in our temperature and metallicity range \nocite{2012MNRAS.427...50L}({Lind}, {Bergemann}, \&
{Asplund} 2012). For the right set of parameters three conditions are fulfilled: The determined Fe
abundance or metallicity is independent of the excitation potential (excitation equilibrium), there
is no trend between Fe abundance and reduced EWs (EW/$\lambda$), and the abundances determined from
Fe\,{\sc i} and Fe\,{\sc ii} lines are equal (ionization equilibrium). The  parameters are found by
minimizing a $\chi^2$ function that is a quadratic form constructed from the conditions mentioned
above. We use the Nelder-Mead or Downhill Simplex algorithm to perform the minimization procedure.
Our implementation follows closely the method described in \nocite{2011RMxAA..47....3S}{Saffe}
(2011).  The error for \Teff was estimated to be the temperature difference that would produce a
difference in Fe abundances of low and high excitation potential lines that is larger than the
standard deviation ($\sigma$) of the determined Fe abundance. In a similar way, the uncertainty for
the microturbulence \vt can be derived as the difference in \vt that produces Fe abundance
differences for different reduced equivalent widths that are larger than $1\sigma$ of the Fe
abundance. The error for the surface gravities is assumed to be the difference in logg that produces
a difference in the Fe abundances derived from Fe\,{\sc i} and Fe\,{\sc ii} lines that is larger
than the standard deviation of the Fe abundance. The error of the metallicity is calculated from the
difference in the Fe\,{\sc i} and Fe\,{\sc ii} abundances combined with the standard deviations for
the Fe abundance determinations for Fe\,{\sc i} and {\sc ii} lines added in quadrature. The
procedure was tested for Fe line EWs measured in a solar spectrum observed with the same setup as
the program stars.  We obtain $\Teff=5756\pm86\,\text{K}$, $\logg=4.41\pm0.11$,
$[\text{Fe/H}]=-0.01\pm0.07$, and $\vt=0.76\pm0.97\,\text{km s$^{-1}$}$. Since the effective
temperature is the most important parameter the determined spectroscopic temperatures were adjusted
by adding the difference of 21\,K compared to the expected solar temperature of 5777\,K. The
spectroscopic parameters are summarized in Table \ref{tab:para_spect}. 
\tabletypesize{\small}
\begin{deluxetable*}{lrrrrrrrr}
\tablecaption{Spectroscopic parameters for the program stars.\label{tab:para_spect}} \tablewidth{0pt}
\tablehead{
\colhead{object} & \colhead{\Teff [K]} &\colhead{$\sigma$} & \colhead{\logg [cm s$^{-2}$]}&
\colhead{$\sigma$} & \colhead{[Fe/H]} &\colhead{$\sigma$} & \colhead{\vt [km s$^{-1}$]}
&\colhead{$\sigma$}
}
\startdata
HIP\,13036 & 4690 & 156 & 1.96 & 0.28 & -0.35 & 0.37 & 1.38 & 0.27 \\ 
HIP\,27380 & 5963 & 69 & 0.68 & 0.20 & -0.24 & 0.44 & 2.70 & 1.30 \\ 
HIP\,30735 & 7269 & 244 & 2.31 & 0.32 & -0.08 & 0.28 & 3.13 & 1.56 \\ 
HIP\,31164 & 5471 & 171 & 3.94 & 0.23 & -0.25 & 0.20 & 1.68 & 0.37 \\ 
HIP\,32063 & 5101 & 83 & 3.30 & 0.21 & -0.19 & 0.16 & 0.87 & 0.16 \\ 
HIP\,32404 & 8192 & 96 & 2.40 & 0.20 & 0.49 & 0.31 & 2.65 & 19.27
\enddata
\tablecomments{Table \ref{tab:para_spect} is published in its entirety in the electronic edition of
\textsl{The Astrophysical Journal}. A portion is shown here for guidance regarding its form and
content.}
\end{deluxetable*}

\subsection{Full spectrum fitting \label{sect:ulyss}}
We determined atmospheric parameters with the full-spectrum fitting \textsl{University of Lyon
Spectroscopic Analysis Software} package ULySS \nocite{2009A&A...501.1269K}({Koleva} {et~al.} 2009).
The observed spectra were fitted against stellar spectra models based on the 3.2 version of the
ELODIE library \nocite{2001yCat.3218....0P}({Prugniel} \& {Soubiran} 2001) represented by an
interpolator \nocite{2011yCat..35319165P}({Prugniel} {et~al.} 2011). The ELODIE library consists of
echelle spectra in the wavelength range $3\,900 - 6\,800\,\AA\,$ with a spectral resolutions of
$\text{R}=42\,000$, containing stars with spectral types from O to M. The interpolator represents a
low resolution ($\text{R}=10\,000$) version of the library and consists of polynomial expansions of
each wavelength element in powers of $\log (\Teff)$, \logg, [Fe/H], and $f(\sigma)$, where
$f(\sigma)$ is a function of the rotational broadening parameterized by the standard deviation of a
Gaussian. The fitting procedure performed by ULySS interpolates a spectrum, convolves it with a
Gaussian and multiplies it with a polynomial to derive the best fit to the observation. The free
parameters of the minimization procedure are the atmospheric parameters \Teff, \logg, and [Fe/H], as
well as two parameters for the Gaussian, the systemic velocity $v_{sys}$ and the dispersion
$\sigma$, and the coefficients of the polynomial. The systemic velocity corrects for uncertainties
in the radial velocities of the stars, the dispersion includes both the instrumental broadening and
the effect of rotation.

\begin{figure}
\centering
\includegraphics[width=\linewidth]{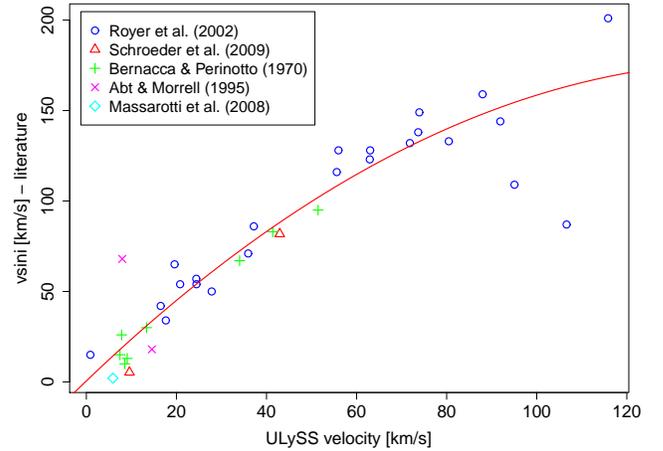}
\caption{Rotational velocities for some of our program stars taken from different literature sources
plotted against the dispersion velocities derived with ULySS. \label{fig:f1}}
\end{figure}

Before the actual fitting, the observed spectra were adapted to the resolution of the interpolated
spectra. We have to ensure that the model has in fact a higher spectral resolution than the
observation \nocite{2011yCat..35319165P}({Prugniel} {et~al.} 2011). This is necessary because it is
convolved with a Gaussian representing a line-of-sight-velocity distribution (LOSVD) during the
analysis, but not always guaranteed because the effective spectral resolution of the spectra is
determined by the rotational broadening and the dispersion of the instrumental broadening. The
observed spectra were therefore convolved with a Gaussian with a fwhm that is higher than necessary
to match the low spectral resolution of the interpolated spectra.  The physical broadening of the
observed spectra due to their rotational velocities can in principle directly be determined from
ULySS if a suitable line-spread function (LSF) is injected that takes into account the varying
relative resolution between the observation and model with wavelength.  However, using an exact LSF
has only a negligible influence on the derived atmospheric parameters
\nocite{2011A&A...525A..71W}({Wu} {et~al.} 2011) and we derived rotational velocities for the
program stars by comparing the measured dispersion velocities with a list of literature values of
$v\sin i$ for our stars from different authors \nocite{2002A&A...393..897R, 2009A&A...493.1099S,
1970CoAsi.239....1B, 1995ApJS...99..135A, 2008AJ....135..209M}({Royer} {et~al.} 2002;
{Schr{\"o}der}, {Reiners}, \&  {Schmitt} 2009; {Bernacca} \& {Perinotto} 1970; {Abt} \& {Morrell}
1995; {Massarotti} {et~al.} 2008). This calibration was restricted to stars with errors in the
dispersion of less than 30\% of the actual value. The result of the calibration can be seen in
Figure \ref{fig:f1}. The solid line shows the fit of a weighted second order polynomial to the
observations with the errors of the dispersion velocities taken as weights. The rotational
velocities are then derived from the dispersion velocities X with the following correlation: $0.556
- 0.008 \cdot \text{X}^2 + 2.386 \cdot \text{X}$.   

The normalized spectra of our program stars were fitted in the wavelength range covered by the
ELODIE spectra. To avoid the introduction of fitting errors caused by the echelle gaps of our
spectra we fitted the wavelength ranges of each order separately. The final temperature for a star
was derived by calculating a sigma-clipped weighted average over the results obtained from all
fitted orders, taking the $\chi^2$ values determined by ULySS as weights. The error for the
temperature is taken as the sigma-clipped and weighted standard deviation. Analogously we derive
values for \logg and [Fe/H] and their errors. 
\tabletypesize{\small}
\begin{deluxetable*}{lrrrrrrrr}
\tablecaption{Parameters for the program stars derived with ULySS.\label{tab:para_ulyss}}
\tablewidth{0pt}
\tablehead{
\colhead{object} & \colhead{\Teff [K]}&\colhead{$\sigma$} & \colhead{\logg [cm s$^{-2}$]}&
\colhead{$\sigma$} & \colhead{[Fe/H]} &\colhead{$\sigma$} & \colhead{disp. vel [km s$^{-1}$]} &
\colhead{$\sigma$}
}
\startdata
HIP\,8066 & 7602 & 162 & 2.90 & 0.71 & 0.12 & 0.10 & 14.21 & 2.03 \\ 
HIP\,13004 & 8250 & 213 & 2.48 & 0.61 & -0.03 & 0.09 & 14.97 & 2.61 \\ 
HIP\,13036 & 5002 &  47 & 2.30 & 0.14 & -0.23 & 0.14 & 6.97 & 1.49 \\ 
HIP\,16001 & 8637 & 152 & 2.10 & 0.34 & -0.10 & 0.08 & 3.37 & 3.37 \\ 
HIP\,20000 & 8693 & 442 & 1.85 & 0.67 & -0.58 & 0.99 & 83.05 & 105.78 \\ 
HIP\,21446 & 7782 & 1559 & 3.06 & 1.70 & -1.25 & 1.12 & 196.88 & 51.19
\enddata
\tablecomments{Table \ref{tab:para_ulyss} is published in its entirety in the electronic edition of
\textsl{The Astrophysical Journal}. A portion is shown here for guidance regarding its form and
content.}
\end{deluxetable*}

Again, atmospheric parameters were derived with a solar spectrum. We obtained
$\Teff=5743\pm73\,\text{K}$, $\logg=4.32\pm0.14$, and $[\text{Fe/H}]=-0.01\pm 0.04$. The difference
of 34\,K compared to the expected solar temperature was added to the derived \Teff for our stars.
The parameters derived with ULySS agree well with the spectroscopic ones. The mean difference in
\Teff is $\sim$20\,K with a $\sigma$ of $\sim$90\,K for stars for which both temperatures estimates
were available. For \logg and [Fe/H] the differences are equally small with $0.25\pm 0.28$ and $0.08
\pm 0.14$. The atmospheric parameters derived with ULySS can be found in Table \ref{tab:para_ulyss}.

\subsection{H-$\alpha$ line fitting}
The H-$\alpha$ line was used to determine effective temperatures for the program stars by fitting
theoretical synthetic spectra to the observations as described in
\nocite{2002A&A...385..951B}{Barklem} {et~al.} (2002).  Problems can arise due to the strong wings
of the line that can extend over a whole spectral order and make the exact determination of the
continuum difficult. Therefore, a special normalization is necessary. The main idea of the procedure
is to determine the continuum of neighboring orders that are not affected by the wings of the line
and to interpolate the continuum for the order containing H-$\alpha$. The procedure is described in
more detail in \nocite{2002A&A...385..951B}{Barklem} {et~al.} (2002). A grid of Barklem's H-$\alpha$
profiles with different temperatures was used and a reduced $\chi^2$ value that quantifies the
difference between observation and synthetic spectrum was derived for each \Teff. The $\chi^2$
function was calculated as the sum of the squared differences between observed and calculated fluxes
divided by the statistical error of the fluxes (1/SNR) for regions of the profile that are not
affected by metal lines. $\chi^2$ values for temperatures in the grid and two different values for
\logg were determined.  The temperature grid ranges from 4\,400 to 7\,500\,K with values for $\logg$
of 3.4 and 3.0. The choice of \logg is influenced by the previously determined spectroscopic value
for \logg but does not have any significant impact on the derived result for \Teff. The best-fit
\Teff is obtained by minimizing the set of $\chi^2$ values and fitting a parabola to the seven
points closest to the minimum of the $\chi^2$ function.  The error of the temperature determination
is given as the error of the fit. The method was applied to a solar spectrum and the parameters for
the normalization procedure were adjusted to get as close to the expected solar temperature as
possible. The systematic error of our procedure was then estimated by determining the solar
temperature with the FTS atlas \nocite{1984sfat.book.....K}({Kurucz} {et~al.} 1984) that was used by
\nocite{2002A&A...385..951B}{Barklem} {et~al.} (2002). For the atlas our method yields a slightly
higher temperature of $\Teff=5799\pm3\,\text{K}$ and the difference of 22\,K to the expected solar
\Teff was subtracted from the H-$\alpha$ temperatures of the program stars.

Unfortunately, for our setup (see Section \ref{sect:observations}) the H-$\alpha$ line is located at
the edge of an order. Therefore, the blue wing of the line can not be used. Also, we restrict the
fit to the upper part of the lines, since the cores are not formed in the photosphere and are
therefore affected by NLTE effects. Due to these difficulties, H-$\alpha$ temperatures could only be
determined for a limited number of stars. The derived temperatures, however,  agree with the
spectroscopic ones with a mean difference of $\sim$30\,K but show a larger standard deviation of
$\sim$200\,K. The temperatures derived by the H-$\alpha$ line fitting are summarized in Table
\ref{tab:para_phot}.

\subsection{Photometric}
Effective temperatures from photometric colors were derived using the calibration of
\nocite{2005ApJ...626..465R}{Ram{\'{\i}}rez} \& {Mel{\'e}ndez} (2005). The intention was to use as
many colors as possible but to consider only photometric data of high quality. The main source for
photometric data used in the analysis was the General Catalogue of Photometric Data (GCPD;
\nocite{1997A&AS..124..349M}{Mermilliod}, {Mermilliod}, \&  {Hauck} 1997). It contains data for many
different photometric systems including \textsl{UBV}, \textsl{uvby}, Vilnius, Geneva, and DDO. In
addition ground-based \textsl{V} magnitudes included in the \textsl{Hipparcos}-Tycho catalog
\nocite{1997yCat.1239....0E}({ESA} 1997) and infrared colors from the Two Micron All Sky Survey
(2MASS), if the observations were not saturated, were used. Some of the program stars have
companions that heavily influence specific photometric colors. In these cases the colors that were
most affected were excluded from the temperature determination. Close binaries and stars with
photometric data of insufficient quality were rejected. Overall, photometric temperatures could be
derived for 142 stars. The average of all available temperatures was adopted as the photometric
\Teff and the standard deviation taken as its error. In cases where only one temperature could be
derived, we assume an error of 100\,K. For all stars the metallicity was assumed to be solar in the
calculations. 

The photometric temperatures are close to the spectroscopic temperatures with a mean difference of
90\,K but show some scatter with a $\sigma$ of 220\,K. The photometric
temperatures are calculated with an estimated reddening (see Section \ref{sect:masses}).

In addition to the temperatures, a photometric calibration was used to calculate \logg values. We
used the derived masses for our stars (see Section  \ref{sect:masses}), the \textsl{V} magnitudes, the
adopted temperatures (see Section \ref{sect:adopted}), and the relation
\begin{align*}
\logg = 0.4(M_\text{bol}+M_{\text{bol}\,\odot}) + \loggsun +
4\log \left( \frac{\Teff}{\Teffsun} \right) + \log \left( \frac{M}{\Msun} \right),
\end{align*}
where $M_\text{bol}=M_\text{V}+BC$. We take $\Teffsun=5780\,\text{K}$, $\loggsun=4.44$, and
$M_{\text{bol}\,\odot}=4.75\,\text{mag}$, as the solar parameters. The photometric \logg agree with
the spectroscopic ones, showing a mean difference of $0.30\pm0.35$. Photometric temperatures and
\logg values are shown in Table \ref{tab:para_phot}.
\tabletypesize{\footnotesize}
\begin{deluxetable}{lrrrrrr}
\tablecaption{H$_\alpha$ and photometric parameters for the program stars. \label{tab:para_phot}} \tablewidth{0pt}
\tablehead{
\colhead{object} & \colhead{\Teff H$_\alpha$ [K]} & \colhead{$\sigma$}& \colhead{\Teff [K]}&
\colhead{$\sigma$} & \colhead{\logg [cm s$^{-2}$]}& \colhead{$\sigma$} 
}
\startdata
HIP\,8066 &  &  & 6997 & 89 & 2.79 & 0.32 \\ 
HIP\,13004 &  &  & 8221 & 100 & 2.74 & 0.35 \\ 
HIP\,19529 &  &  & 7570 & 201 & 3.10 & 0.26 \\ 
HIP\,19823 &  &  & 6111 & 206 & 2.39 & 0.15 \\ 
HIP\,21084 &  &  & 7619 & 100 & 3.09 & 0.36 \\ 
HIP\,21446 &  &  & 7532 & 547 & 3.00 & 0.20
\enddata
\tablecomments{Table \ref{tab:para_phot} is published in its entirety in the electronic edition
of \textsl{The Astrophysical Journal}. A portion is shown here for guidance regarding its form and
content.}
\end{deluxetable}

\subsection{Adopted parameters \label{sect:adopted}}
\begin{figure}
\centering
\includegraphics[width=\linewidth]{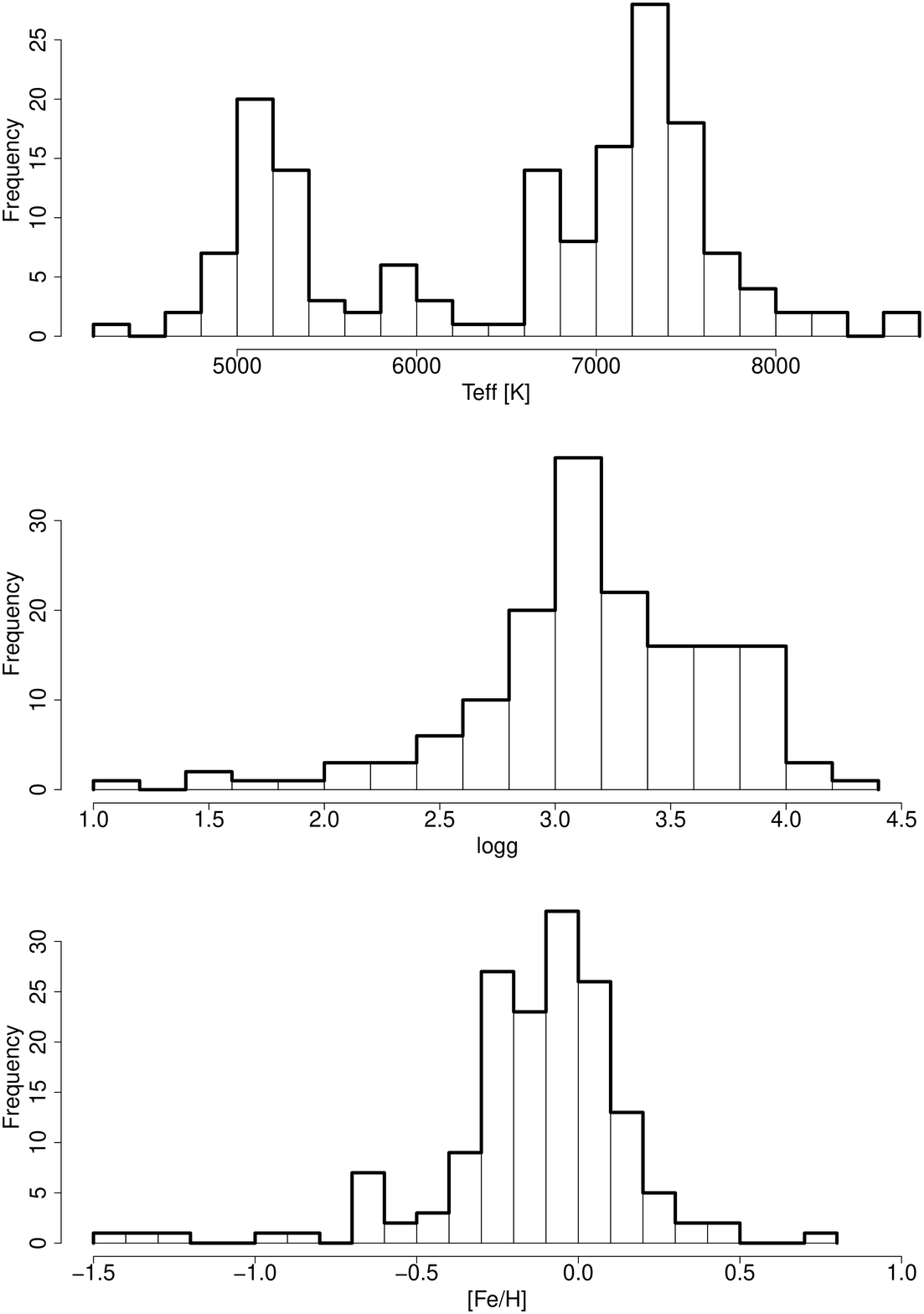}
\caption{Histograms of the adopted parameters of our program stars. \label{fig:f2}}
\end{figure}

We adopted the mean of all available temperature determinations as the effective temperature for our
program stars and take the standard deviation as the error. The number of available temperatures
varies and mainly depends on the rotational broadening of the lines in the spectrum. In almost all
cases, however, a photometric and ULySS temperature can be used. If only one temperature is
available the error of the individual temperature determination was also taken as the uncertainty
for the final adopted temperature. Note that the errors in the individual temperatures derived with
the full spectrum fitting method (Section \ref{sect:ulyss}) for stars that are fast rotators can be
quite large. The agreement with the photometric temperatures for objects where at least two
temperatures could be derived, however, justifies the use of the ULySS temperatures also in cases
where it is the only temperature indicator available.  Furthermore, we include the uncertainty in
temperature in the error determination of the derived elemental abundances (see Section
\ref{sect:abund}).  For the surface gravities the mean and standard deviation of the values from the
spectroscopic and ULySS determination was taken as the adopted value and its error, whenever
possible. If only one of these \logg could be determined again the error of this determination was
taken as the adopted error. For stars with only photometric temperatures the photometric \logg and
the associated error was taken. Similarly, the mean of the metallicities derived from the
spectroscopic and ULySS determination was taken as the adopted [Fe/H] and the standard deviation as
its error. A solar metallicity was assumed for stars with only photometrically determined \Teff and
\logg. An exception was made for HIP\,95497 (RR\,Lyr) for which we take a metallicity of -1.45\,dex.
This is the mean of the two metallicities from \nocite{2010A&A...519A..64K}{Kolenberg} {et~al.}
(2010) and \nocite{1999PASJ...51..961T}{Takeda} {et~al.} (1999), see Section \ref{sect:lit_comp}.
The microturbulence and its error were taken from the spectroscopic determination. For stars with no
derived microturbulence we assume $\vt=1.8\pm0.35\, \text{km s}^{-1}$. Furthermore, the rotational
velocities from our calibration described in section \ref{sect:ulyss} were adopted. All adopted
atmospheric parameters can be found in Table \ref{tab:para_adop} and histograms of the adopted
values for \Teff, \logg, and [Fe/H] are shown in Figure \ref{fig:f2}.
\tabletypesize{\small}
\begin{deluxetable*}{lrrrrrrrrrrr}
\tablecaption{Adopted atmospheric parameters for the program stars. The fourth column shows the number of
different temperature indicators that were used in the determination of \Teff. \label{tab:para_adop}}
\tablewidth{0pt}
\tablehead{
\colhead{object} & \colhead{\Teff [K]} & \colhead{$\sigma$} & \colhead{n} & \colhead{\logg [cm
s$^{-2}$]}& \colhead{$\sigma$} & \colhead{[Fe/H]} & \colhead{$\sigma$} & \colhead{ \vt [km
s$^{-1}$]}& \colhead{$\sigma$} & \colhead{$v_{\sin i}$ [km s$^{-1}$]} & \colhead{$\sigma$}
}
\startdata
HIP\,8066 & 7300 & 428 &   2 & 2.90 & 0.71 & 0.12 & 0.10 & 1.80 & 0.35 & 32.84 & 2.03 \\ 
HIP\,13004 & 8236 &  21 &   2 & 2.48 & 0.61 & -0.03 & 0.09 & 1.80 & 0.35 & 34.47 & 2.61 \\ 
HIP\,13036 & 4846 & 221 &   2 & 2.13 & 0.24 & -0.29 & 0.08 & 1.38 & 0.27 & 16.80 & 1.49 \\ 
HIP\,16001 & 8637 & 152 &   1 & 2.10 & 0.34 & -0.10 & 0.08 & 1.80 & 0.35 & 8.51 & 3.37 \\ 
HIP\,19529 & 7570 & 201 &   1 & 3.10 & 0.26 & 0.00 & 0.25 & 1.80 & 0.35 & 89.72 & 109.68 \\ 
HIP\,19823 & 6111 & 206 &   1 & 2.38 & 0.15 & 0.00 & 0.25 & 1.80 & 0.35 & 10.69 & 6.35
\enddata
\tablecomments{Table \ref{tab:para_adop} is published in its entirety in the electronic edition of
\textsl{The Astrophysical Journal}. A portion is shown here for guidance regarding its form and
content.}
\end{deluxetable*}
The first panel clearly shows the separation of the sample into two groups with temperatures below
and above $\sim 6500\,\text{K}$. This separation illustrates the limited number of stars than
can be found in the center of the Hertzsprung gap. 

\subsection{Reddening, Mass, and Luminosity \label{sect:masses}}
The reddening of our stars was estimated according to the procedure described in
\nocite{2007AJ....133.2464L}{Luck} \& {Heiter} (2007). First, the reddening of the stars was
calculated with the EXTINCT code from \nocite{1997AJ....114.2043H}{Hakkila} {et~al.} (1997), using
the position of the objects and the distances calculated from the parallaxes of the new reduction of
the \textsl{Hipparcos} catalog \nocite{2007A&A...474..653V}({van Leeuwen} 2007). Then, the reddening
for the reddening-free Local Bubble, i.e. all reddening calculated for a distance of 75\,pc, was
subtracted. 

\begin{figure}
\centering
\includegraphics[width=\linewidth]{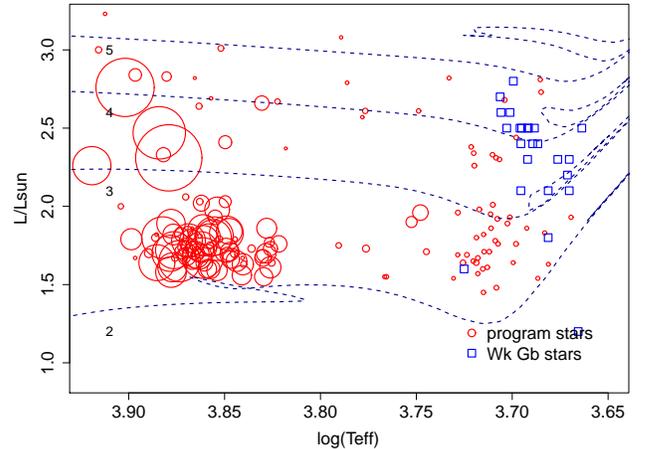}
\caption{Program stars in the HRD. The circle size correlates with the rotational velocities of the
objects. Evolutionary tracks, labeled with the respective stellar mass in \Msun, were calculated by
\protect\nocite{2008A&A...484..815B, 2009A&A...508..355B}{Bertelli} {et~al.} (2008, 2009) for a
solar composition ($\text{Z}_\odot=0.017$, $\text{Y}_\odot=0.260$). \label{fig:f3}}
\end{figure}

Luminosities for the program stars were calculated with the \textsl{Hipparcos} parallaxes and the
GCPD visual magnitudes and their errors, whenever possible. For stars with no GCPD V magnitudes,
\textsl{Hipparcos} magnitudes and their errors were used. We take the adopted temperatures and
applied a bolometric correction according to \nocite{1999A&AS..140..261A}{Alonso} {et~al.} (1999).
With the luminosities and adopted temperatures, the program stars can be placed in the HRD. This is
shown in Figure \ref{fig:f3}.

The masses of the stars were determined by interpolating between the evolutionary tracks from
\nocite{2008A&A...484..815B, 2009A&A...508..355B}{Bertelli} {et~al.} (2008, 2009). We rely on the
evolutionary tracks for a solar composition ($\text{Z}_\odot=0.017$, $\text{Y}_\odot=0.260$).
Changing the metallicity does not significantly affect the masses of our objects. The masses of the
program stars are almost exclusively in the range $2-5\,\Msun$ with a mean mass of
$2.51\pm0.76\,\Msun$. To check the influence of the chosen evolutionary tracks, masses were also
derived from evolutionary tracks from \nocite{2004ApJS..155..667D}{Demarque} {et~al.} (2004). These
masses were higher by $0.31\pm0.21\,\Msun$. The lower number of objects above $3\,\Msun$ in our
sample is due to the constraints from the target selection and the generally low population of stars
in this area of the HRD in the \textsl{Hipparcos} catalog. The results for the mass and luminosity
determination are summarized in Table \ref{tab:masses}.
\tabletypesize{\small}
\begin{deluxetable*}{lrrrrrrrr}
\tablecaption{Masses, distances, and luminosities of the program stars. \label{tab:masses}}
\tablewidth{0pt}
\tablehead{
\colhead{object} & \colhead{plx [mas]}& \colhead{$\sigma$} & \colhead{d [pc]} & \colhead{$\sigma$} &
\colhead{$\log$ L} & \colhead{$\sigma$} & \colhead{M [\Msun]} & \colhead{$\sigma$} 
}
\startdata
HIP\,8066 & 1.97 & 0.66 & 508 & 170 & 2.64 & 0.29 & 3.61 & 0.68 \\ 
HIP\,13004 & 1.48 & 0.58 & 676 & 265 & 3.00 & 0.34 & 4.44 & 1.00 \\ 
HIP\,13036 & 2.16 & 0.56 & 463 & 120 & 2.81 & 0.23 & 4.58 & 0.68 \\ 
HIP\,16001 & 1.33 & 0.52 & 752 & 294 & 2.85 & 0.34 & 4.01 & 0.88 \\ 
HIP\,19529 & 2.62 & 0.71 & 382 & 103 & 2.31 & 0.24 & 2.91 & 0.43 \\ 
HIP\,19823 & 2.33 & 0.33 & 429 & 61 & 2.79 & 0.12 & 4.10 & 0.32
\enddata
\tablecomments{Table \ref{tab:masses} is published in its entirety in the electronic edition of
\textsl{The Astrophysical Journal}. A portion is shown here for guidance regarding its form and
content.}
\end{deluxetable*}

\section{Abundances} \label{sect:abund}
The abundance analysis focuses on the elements that are key constituents of the CNO-cycle and are
altered in giant stars after the first dredge-up: C and O. The focus on these
elements is also founded on the number of spectral lines that are accessible for the two elements.
Unfortunately, the determination of a N abundance is prevented by the lack of usable N lines
in our spectra. For the abundance analysis described below the EWs of suitable lines were measured
with the IRAF routine splot. The lines were de-blended when necessary and corrected for the
influence of other elements. The abundances were then determined with MOOG using a model atmosphere
with the adopted parameters as input. In addition to the program stars the EWs of lines in a solar
spectrum observed with the same setup were measured and a solar abundance for each line was derived.
The solar abundances were used to determine relative abundances for our elements of interest.

\subsection{Carbon \label{sect:carbon}}
We use three C\,{\sc i} lines at 5052, 5380, and 6587\,\AA\, as the main indicators for the C
abundance. The 5052\,\AA \, line is close to an Fe\,{\sc i} and a Cr\,{\sc i} line and has to be
carefully de-blended. The signal to noise ratio of our spectra is sufficiently high but difficulties
are caused by the high rotational velocities in some stars. The line at 6587\,\AA\, is very weak but
can nevertheless be used to derive abundances for a large fraction of our objects. For all three
lines the excitation potentials and \loggf values as published in
\nocite{2005A&A...431..693A}{Asplund} {et~al.} (2005) were taken. With the observed solar spectrum
and solar parameters of $\Teffsun=5780\,\text{K}$, $\loggsun=4.44$, the C abundances of 8.39
\footnote{Abundances are given as $\log \epsilon (\text{X}) = \log(N_\text{X}/N_\text{H}) + 12$},
8.41, and 8.32 for the lines at 5052, 5380, and 6587\,\AA\, were derived. 

We use the C\,{\sc i} line at 8335\,\AA\, as an additional abundance indicator.  However, since this
line can be severely blended with a telluric component it could not be used for the majority of our
objects. In two cases, for HIP\,81933 and HIP\,82093, the 8335\,\AA\, line is the only indicator for
the C abundance. For this line a solar abundance of 8.29 was obtained. Again, the excitation
potential and \loggf value from \nocite{2005A&A...431..693A}{Asplund} {et~al.} (2005) were used. 

The C\,{\sc i} triplet at 9078, 9088, and 9094\,\AA\, is present in our spectra but strongly blended
by telluric lines that make it impossible to derive accurate abundances from them. The C\,{\sc i}
line at 9111\,\AA\, is strong enough in our spectra. However, the line is close to a Fe\,{\sc i}
line and in many cases blended. Furthermore, it suffers from severe NLTE effects
\nocite{2001A&A...379..936P}({Przybilla}, {Butler}, \&  {Kudritzki} 2001) and was therefore not used
for an abundance determination. The atomic data for the lines used for the C abundance determination
are presented in Table \ref{tab:lines}. 

The C abundance is derived as the mean from the abundances of the individual lines and is shown in
Table \ref{tab:Cabund}. The error is given as the standard deviation. In cases where only the
abundance of one line was available this error is assumed to be 0.1\,dex. Overall, C abundances for
66 stars could be derived with the described method. The other stars in the sample show strongly
broadened lines that prevent the measurement of individual EWs. We estimated the effect of a change
in temperature on the C abundances. Changing \Teff by 100\,K leads to a change in the C abundance of
less than 0.1\,dex.  Using the derived uncertainties for the adopted temperatures an additional
temperature dependent error was calculated for the C abundance of each star and included in the
determination of the error of [C/H]. Luck \& Heiter (2006, Figure 3) showed that the C abundances
from our three C\,{\sc i} lines gave very similar results for stars with $\Teff \geq 5250\,$K but at
lower temperatures the 5380\,\AA\, line gave higher abundances presumably due to an unknown blend
and the 5052\,\AA\, line gave lower abundances due to interference from the wing of a Fe\,{\sc i}
line. Few of our stars (Figure \ref{fig:f2}) have temperatures cooler than 5250\,K and there are
no significant differences in the abundances derived from the 5380 and 5052\,\AA\, line in these
cases.  For the cool stars, the primary C abundance indicator chosen by Luck \& Heiter was a C$_2$
Swan band feature at 5135\,\AA. This feature is present in our coolest stars but disappears for
hotter temperatures. 

We used the stars with C abundances derived from the individual lines to infer C abundances for the
rest of the stars in our sample that exhibit high velocity broadening. For this purpose the strength
of the CH G band was measured for each star by integrating the flux over the wavelength range
$4290-4320\,\AA$. Then, a weighted multivariate linear model was calculated that sets the C abundance
in relation to the measured strength of the G band and the atmospheric parameters \Teff, \logg, and
[Fe/H], taking the previously determined errors in the C abundances as weights. The model was
trained with the sample of stars for which both C abundances from the individual lines and the
strength of the CH G band could be determined. The Am/Ap stars (see Section \ref{sect:outliers}) in
the sample were excluded in the calculation of the model. The mean residual of the resulting model
is $-0.01\pm 0.31$\,dex. The model was used to predict the C abundances for the stars with high
rotational velocities. Finally, a metallicity dependent correction was added to adjust the trend
with [Fe/H] and account for the fact that many of the high rotators have high metallicities that
extend beyond the range of the training set. The strength of the G band depends on \Teff and gets
weaker for increasing temperatures. Due to the high rotation in the hottest stars in our sample it
is impossible to visually detect if the absorption in that region is due to CH or other metal lines.
A comparison with synthetic spectra calculated with varying C abundances in our parameters range,
however, suggests that the contribution of CH gets negligible for stars hotter than 7500\,K and we
restrict our G band abundance determination to temperatures cooler than that. The error for the
abundances derived in this way is estimated to be 0.32\,dex and consists of the standard deviation
of the residuals and the mean error of the C abundances from the training set added in quadrature.
The C abundances derived in this way are also shown in Table \ref{tab:Cabund}. For the C abundances
derived from the individual lines [C/H] is the mean of the abundances relative to the solar
abundances of each line. For the C abundances determined with the G band [C/H] is the calibrated C
abundance relative to the mean solar C abundance (8.35).
\begin{deluxetable}{lrrrrl}
\tablecaption{C abundances for the program stars. The last column indicates the method used for the
abundance determination (l = lines, G = CH G band model). See text for details. \label{tab:Cabund}}
\tablewidth{0pt}
\tablehead{
\colhead{object} & \colhead{C} & \colhead{$\sigma$} & \colhead{[C/H]} & \colhead{$\sigma$} & 
\colhead{method} 
}
\startdata
HIP\,8066 & 8.26 & 0.32 & -0.09 & 0.53 & G \\ 
HIP\,13036 & 7.82 & 0.39 & -0.58 & 0.43 & l \\ 
HIP\,16001 & 8.75 & 0.03 & 0.37 & 0.15 & l \\ 
HIP\,19823 & 7.84 & 0.13 & -0.54 & 0.27 & l \\ 
HIP\,27380 & 7.70 & 0.09 & -0.67 & 0.19 & l \\ 
HIP\,27747 & 8.11 & 0.32 & -0.24 & 0.43 & G
\enddata
\tablecomments{Table \ref{tab:Cabund} is published in its entirety in the electronic edition of
\textsl{The Astrophysical Journal}. A portion is shown here for guidance regarding its form and
content.}
\end{deluxetable}
\begin{figure}
\centering
\includegraphics[width=\linewidth]{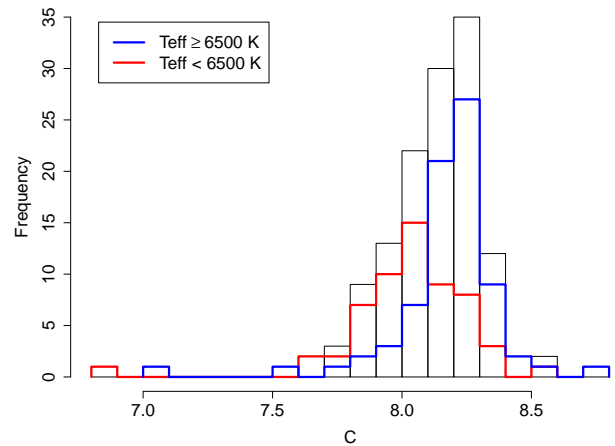}
\caption{C abundances for our program stars. \label{fig:f4}}
\end{figure}

In Figure \ref{fig:f4} a histogram of the C abundances derived with both methods is shown. Stars
with temperatures below 6500\,K show in general smaller C abundances than the group of stars with
higher temperatures.  

\begin{figure}
\centering
\includegraphics[width=\linewidth]{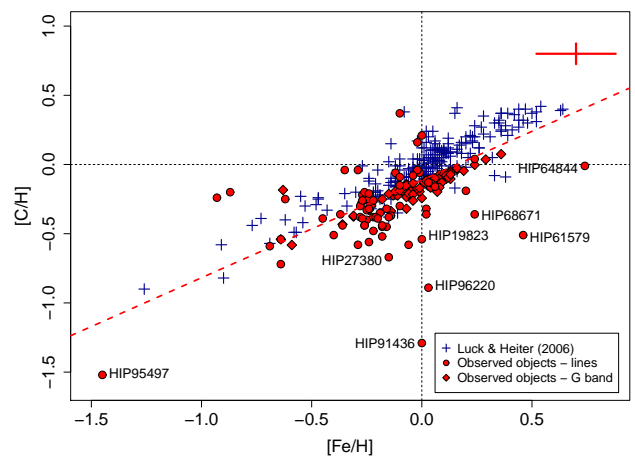}
\caption{Carbon abundance relative to solar for our program stars. The cross indicates the typical
error of the abundances derived with the individual lines. The dashed line is a linear fit to the
stars from \protect\nocite{2006AJ....131.3069L}{Luck} \& {Heiter} (2006) shifted by an offset of
-0.1\,dex in [C/H], see text for details. \label{fig:f5}}
\end{figure}

In Figure \ref{fig:f5} the C abundances are compared to data for nearby dwarfs of spectral types F-K
from \nocite{2006AJ....131.3069L}{Luck} \& {Heiter} (2006), covering a \Teff range from $4100 -
7141\,\text{K}$. Luck \& Heiter derived their abundances by synthesizing the C\,{\sc i} 5380 and
6578\,\AA\, lines and a feature of the C$_2$ Swan system at 5135\,\AA. 

The offset of our [C/H] -- [Fe/H] trend from that defined by Luck \& Heiter's F-K dwarfs arises
presumably from systematic differences in the [C/H] and/or the [Fe/H] abundances: a difference of
-0.1\,dex in [C/H] (see Figure \ref{fig:f5}) is shown to reconcile the two trends but a similar
adjustment to [Fe/H] achieves a comparable reconciliation. Unfortunately, the two samples do not have
stars in common so that direct determinations of the systematic differences are not possible.
Differences at the level of 0.1\,dex seem inevitable when analyses of different samples are
undertaken by different groups: different lines with different $gf$-values and different families of
model atmospheres, different basic defining parameters are some of the key factors.

Luck \& Heiter compare their [Fe/H] measurements with results from eight different analyses by other
authors and find mean differences from -0.01 to +0.07\,dex in considering large numbers of common
stars. For [C/H], they find a mean difference of -0.07\,dex from 62 stars in common with
\nocite{2004A&A...426..619E}{Ecuvillon} {et~al.} (2004) where the difference is +0.1\,dex with our
sample. Our sample is systematically warmer than Luck \& Heiter's and, thus, non-LTE effects may
introduce an offset between the C and/or the Fe abundances. The lines used in our analysis are
generally not known to be severely affected by non-LTE effects but small corrections in the order of
$\approx 0.04$\,dex were found by \nocite{2005A&A...431..693A}{Asplund} {et~al.} (2005) for 3 of the
lines considered here. In light of the fact that only large [C/H]
differences would be an indicator for a possible connection between the sample stars and peculiar stars on
the giant branch like the wk Gb stars, the offset of 0.1 dex is irritating
but not fatal. We therefore refrained from performing a detailed analysis of possible non-LTE
effects. Moreover, the comparable comparison for [O/H] -- [Fe/H] (see Figure \ref{fig:f7})
shows no offset between our and Luck \& Heiter's samples.

\subsection{Oxygen \label{sect:oxygen}}
We employ the [O\,{\sc i}] lines at 6300, 6363\,\AA, and O\,{\sc i} at 5577\,\AA, and the O\,{\sc i}
triplet at 7771, 7774, and 7775\,\AA. The [O\,{\sc i}] line at 6300\,\AA\, is blended with a
Ni\,{\sc i} line at almost the same position. To correct for the Ni influence the expected EW for
the Ni line was calculated with our adopted atmospheric parameters and the \loggf value from
\nocite{2001ApJ...556L..63A}{Allende Prieto}, {Lambert}, \&  {Asplund} (2001). Then the Ni EW was
subtracted from the measured EWs of the O line before the O abundance was determined. The excitation
potential and \loggf value for the O line were taken from \nocite{2001ApJ...556L..63A}{Allende
Prieto} {et~al.} (2001). For the solar spectrum an abundance of 8.66 was derived for this line. The
6363\,\AA\, line is blended by CN lines and located in an Ca\,{\sc i} auto-ionization feature. While
the influence of the CN lines is difficult to estimate, the effect of the Ca feature can be
neglected by adjusting the continuum at the EW line determination accordingly.  The line is barely
visible in the solar spectrum and the measured equivalent width very small. For this line and for
the [O\,{\sc i}] 5577\,\AA\, with a similarly low EW we therefore assumed a solar abundance of 8.69
in accordance to \nocite{2009ARA&A..47..481A}{Asplund} {et~al.} (2009). The O\,{\sc i} triplet lines
are known to be affected by NLTE effects. For each star non-LTE corrections as described in
\nocite{2007A&A...465..271R}{Ram{\'{\i}}rez}, {Allende Prieto}, \&  {Lambert} (2007) were calculated
for each line separately.  The non-LTE corrected solar abundances are 8.75, 8.74, and 8.72 for the
lines at 7771, 7774, and 7775\,\AA. The final O abundance and its error are derived as the mean and
standard deviation of the O abundances determined from the individual lines. For stars with only one
O abundance indicator the error is assumed to be 0.15\,dex. The atomic data used for the O lines is
again summarized in Table \ref{tab:lines}.  The O abundances of our program stars can be found in
Table \ref{tab:Oabund}.
\tabletypesize{\small}
\begin{deluxetable}{llrrrl}
\tablecaption{Lines used for the abundance determination. \label{tab:lines}} \tablewidth{0pt}
\tablehead{
\colhead{line [\AA]}& \colhead{element} & \colhead{$\chi$ [eV]} & \colhead{$\loggf$} &
\colhead{source}
}
\startdata
5052.167 & C\,{\sc i} & 7.685 & -1.304 & 1\\
5380.340 & C\,{\sc i} & 7.685 & -1.615 & 1\\
6587.610 & C\,{\sc i} & 8.537 & -1.021 & 1\\
8335.150 & C\,{\sc i} & 7.680 & -0.440 & 1\\
5577.340 & [O\,{\sc i}] & 1.970 & -8.240 & 4\\
6300.304 & [O\,{\sc i}] & 0.000 & -9.717 & 3\\
6363.790 & [O\,{\sc i}] & 0.020 &-10.185 & 2\\
7771.944 & O\,{\sc i} & 9.146 & 0.369 & 2\\
7774.166 & O\,{\sc i} & 9.146 & 0.223 & 2\\
7775.388 & O\,{\sc i} & 9.146 & 0.002 & 2
\enddata
\tablecomments{Source: 1: \nocite{2005A&A...431..693A}{Asplund} {et~al.} (2005),
2: \nocite{2004A&A...417..751A}{Asplund} {et~al.} (2004),
3: \nocite{2001ApJ...556L..63A}{Allende Prieto} {et~al.} (2001),
4: \nocite{2011ApJ...743..135R}{Ram{\'{\i}}rez} \& {Allende Prieto} (2011)}
\end{deluxetable}
\tabletypesize{\small}
\begin{deluxetable}{lrrrrl}
\tablecaption{O abundances for the program stars. The last column indicates the method used for the
abundance determination (l = lines, tr = O\,{\sc i} triplet model). See text for details.
\label{tab:Oabund}} \tablewidth{0pt}
\tablehead{
\colhead{object} & \colhead{O} & \colhead{$\sigma$} & \colhead{[O/H]} & \colhead{$\sigma$} & 
\colhead{method} 
}
\startdata
HIP\,8066 & 8.99 & 0.17 & 0.28 & 0.46 & tr \\ 
HIP\,13004 & 9.16 & 0.17 & 0.45 & 0.17 & tr \\ 
HIP\,13036 & 8.76 & 0.23 & 0.05 & 0.30 & l \\ 
HIP\,16001 & 9.33 & 0.17 & 0.62 & 0.23 & tr \\ 
HIP\,19529 & 9.20 & 0.17 & 0.49 & 0.26 & tr \\ 
HIP\,19823 & 8.90 & 0.17 & 0.19 & 0.27 & tr
\enddata
\tablecomments{Table \ref{tab:Oabund} is published in its entirety in the electronic edition of
\textsl{The Astrophysical Journal}. A portion is shown here for guidance regarding its form and
content.}
\end{deluxetable}
\begin{figure}
\centering
\includegraphics[width=\linewidth]{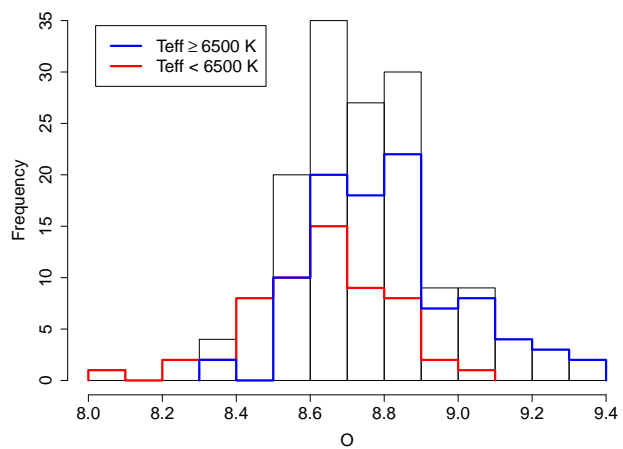}
\caption{O abundances for our program stars. \label{fig:f6}}
\end{figure}

The abundances derived from the different lines agree well. Taking the [O\,{\sc i}] line at
6300\,\AA\, as a reference the [O/H] abundances derived from the other lines show a mean differences
of $-0.07\pm 0.19\,\text{dex}$ (5577\,\AA), and $-0.05\pm0.16\,\text{dex}$ (6363\,\AA), where the
[O/H] abundances for each line are relative to the solar abundance derived with the same line. The O
abundances of the triplet lines show larger differences for strong lines that suffer from
saturation. We therefore consider only abundances derived from triplet lines with $\text{EWs} <
200\,\text{m\AA}$. With this constraint the difference reduces to $0.00\pm0.19\,\text{dex}$. The
scatter in the abundances indicated by the standard deviation is due to the higher temperature
sensitivity of the triplet lines. Adjusting the temperatures of the objects by $53\pm136\,\text{K}$
would result in identical O abundances for both the 6300\,\AA\, line and the O triplet.

The O abundances for stars with high rotational velocities were derived in a way similar to the
derivation of the C abundances. The flux was integrated for all stars in the wavelength range around
the O\,{\sc i} triplet (7765 - 7780\,\AA). This region contains more relevant contribution from O
absorption than the 6300\,\AA\,line. Then, a linear model was constructed from the stars with
non-LTE corrected O abundances derived from the O triplet lines using the integrated flux and the
atmospheric parameters, and used to predict the O abundances for the high rotation stars taking
again the errors in the O abundances as weights. The mean residual of this model is
$0.01\pm0.09$\,dex. The error for the O abundances derived in this way is 0.17\,dex. The calculated
abundances for the high rotation stars are displayed in Table \ref{tab:Oabund} and a histogram of
the distribution of the O abundances is shown in Figure \ref{fig:f6}. As for C the high rotators,
with their higher metallicities, show slightly higher O abundances.

\begin{figure}
\centering
\includegraphics[width=\linewidth]{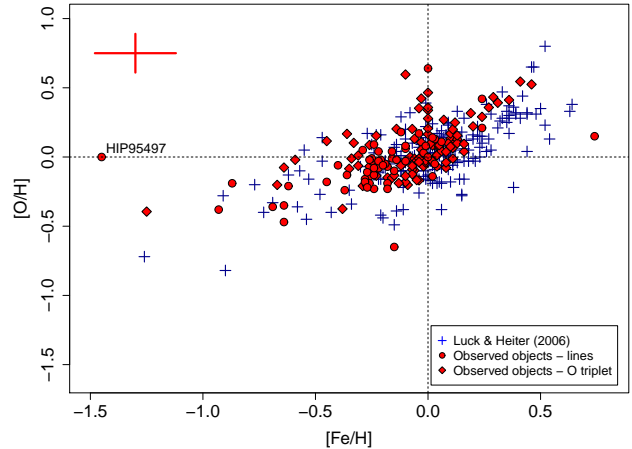}
\caption{Oxygen abundance relative to solar for our program stars. The cross indicates the typical
error of the abundances derived with the individual lines. \label{fig:f7}}
\end{figure}

In Figure \ref{fig:f7} the derived O abundances are compared with the results from
\nocite{2006AJ....131.3069L}{Luck} \& {Heiter} (2006). They use a synthesis of the forbidden line at
6300\,\AA\, as the sole indicator for the O abundance. While Luck \& Heiter use the same \loggf
value for the O line as we do, the strength of the blending Ni line is determined under the
assumption that [Ni/Fe] = 0 with the experimental \loggf value from
\nocite{2003ApJ...584L.107J}{Johansson} {et~al.} (2003).  

Overall, the distribution of O abundances at a given [Fe/H] are slightly more dispersed than the C
abundances. Luck \& Heiter attributed the larger scatter of the O abundances compared to C partly to
the difficulty of determining the O abundance from the weak [O\,{\sc i}] line.  The [O/H] abundances
agree well despite the differences in the lines used for the analysis: a slight offset exists with
our program stars being about 0.06\,dex higher than the linear fit to Luck \& Heiter's data.
\nocite{2006A&A...445..633E}{Ecuvillon} {et~al.} (2006) derived O abundances for F-K stars including
many hosting exoplanets.  Their abundance indicators were the [O\,{\sc i}] 6300\,\AA\, lines, the
O\,{\sc i} 7774\,\AA\, triplet and several ultraviolet OH lines. Luck \& Heiter compared their O
abundances for 48 stars in common with Ecuvillon et al. and found a mean difference of only
-0.02\,dex.  RR\,Lyr (HIP\,95497) shows an exceptional high O abundance that is much higher than
found in previous determinations (see Section \ref{sect:lit_comp}).

\subsection{Comparison with literature values \label{sect:lit_comp}}
\tabletypesize{\small}
\begin{deluxetable*}{llllrlrlrl}
\tablecaption{Comparison of derived abundances of C, O, and Fe of our program stars with literature
abundances. The rows with no source specified refer to our analysis. Asterisks mark mean values of
pulsating stars. \label{tab:literature}} \tablewidth{0pt}
\tablehead{
\colhead{object} & \colhead{\Teff} & \colhead{\logg} & \colhead{Fe} & \colhead{[Fe/H]}&
\colhead{C} & \colhead{[C/H]} & \colhead{O} & \colhead{[O/H]} & \colhead{source}
}
\startdata
HIP\,51056 & 7120&3.34& 7.51 &      &8.40 &      & 8.76 &      & {\nocite{2003ARep...47..865R}{Rachkovskaya} (2003)}\\
           & 7206&3.54& 7.61 &      &8.29 &      & 8.80 &      & \\[1mm]
HIP\,64844 & 7300&3.20&      & 0.45 &     &-0.12 &      & 0.62 & {\nocite{2012ARep...56..850G}{Galeev} {et~al.} (2012)}\\
           & 7534&3.68&      & 0.57 &     &-0.46 &      &      & {\nocite{1995ApJ...451..747R}{Russell} (1995)}\\
           & 7361&2.97&      & 0.74 &     &-0.01 &      & 0.15 & \\[1mm]
HIP\,70602 & 7780&3.58& 7.58 &      &     &      & 8.42 &      & {\nocite{1999PASJ...51..961T}{Takeda} {et~al.} (1999)}\\
           & 7514&3.13& 7.81 &      &     &      & 9.13 &      & \\[1mm]
HIP\,87998 & 6483&2.35& 7.00 &      &8.38 &      & 8.58 &      & {\nocite{2008AN....329....4A}{Adelman} {et~al.} (2008)}\\
           & 6575&2.30& 7.10 &      &8.14 &      & 8.56 &      & {\nocite{1995AJ....110.2425L}{Luck} \& {Wepfer} (1995)}\\
           & 7111&3.66& 7.46 &      &8.22 &      & 8.70 &      & \\[1mm]
HIP\,95497 & 6125&2.40&      &-1.57 &7.16 &      & 8.02 &      & {\nocite{2010A&A...519A..64K}{Kolenberg} {et~al.} (2010)}\\
           & 6622&2.31&      &-1.32 &     &      & 7.83 &      & *{\nocite{2006PASJ...58..389T}{Takeda} {et~al.} (2006)}\\
           & 6304&3.04&      &-1.45 &6.88 &      & 8.72 &      & \\[1mm]
HIP\,96302 & 5300&2.80&      & 0.10 &     &-0.18 &      &-0.22 & {\nocite{2008ARep...52..226B}{Balega} {et~al.} (2008)}\\
           & 5132&3.04&      &-0.26 &     &-0.44 &      &-0.02 & \\[1mm]
HIP\,97985 & 5450&1.79&      &-0.67 &     &-0.22 &      &      & {\nocite{2008PASJ...60..781T}{Takeda}, {Sato}, \&  {Murata} (2008)}\\
           & 5875&2.00& 7.10 &      &8.50 &      & 8.90 &      & {\nocite{1999PASP..111...84V}{Vanture} \& {Wallerstein} (1999)}\\
           & 5407&2.19& 6.63 &-0.87 &8.21 &-0.20 & 8.47 &      & \\[1mm]
HIP\,104185& 6245&2.43&      & 0.08 &     &-0.19 &      & 0.05 & *{\nocite{2013MNRAS.432..769T}{Takeda} {et~al.} (2013)}\\
           & 6170&2.35&      & 0.10 &     &-0.05 &      & 0.05 & {\nocite{2008AJ....136...98L}{Luck} {et~al.} (2008)}\\
           & 6267&2.43&      & 0.11 &     &-0.12 &      & 0.01 & {\nocite{2002A&A...381...32A}{Andrievsky} {et~al.} (2002)}\\
           & 6155&1.42&      & 0.02 &     &-0.36 &      &-0.14 &
\enddata
\end{deluxetable*}

Few of our stars have been previously analyzed.  Table \ref{tab:literature} shows a comparison of
the published abundances.  Naturally, this suffers from different techniques and data used. We
therefore directly quote the parameters and abundances as presented by the authors. In general
literature and newly determined abundances agree within the uncertainties of the determination.
However, there are some noteworthy discrepancies.  

The most significant deviation is the O abundance of RR\,Lyr (HIP\,95497).
\nocite{2010A&A...519A..64K}{Kolenberg} {et~al.} (2010) derived an O abundance of 8.02 by spectral
synthesis of 6 O features.  Their result agrees well with \nocite{2006PASJ...58..389T}{Takeda}
{et~al.} (2006), who determined an O abundances of 7.96 by synthesizing the wavelength regions of
6154.5 - 6159.5\AA\, and 7770-7777\AA. Both quoted O abundances are the mean of the abundances
derived for several pulsation phases. Due to the blending of telluric O$_2$ lines, Takeda et al.
made use of the [O\,{\sc i}] forbidden line in only one case, for which they measured an EW of
3.9\,m\AA\, (O = 7.76). This compares to an EW of 23.5\,m\AA\, in our spectra (O = 8.97). The
difference in the measured EWs might suggest a residual telluric component in our O line but all
lines were carefully de-blended when the EWs were measured. The non-LTE corrected O abundance
derived by the O triplet in our spectra is 8.64 a value in agreement with the [O\,{\sc i}] line but
higher than the one determined by Takeda et al. Despite the differences in the determined abundances
it is interesting that HIP\,95497 appears in the sample at all. The prototype of the RR\,Lyrae stars
is a horizontal branch object and has therefore already experienced a first ascent of the giant
branch.  Unfortunately, with the given target selection and analysis method it is not possible to
distinguish horizontal branch stars from late MS and subgiant stars. 

For HIP\,104185 the derived \logg is lower than in previous determinations. This leads to both lower
abundances in O and C. 

For HIP\,97985 there appears to be a larger difference between our O abundance and the one
determined by \nocite{1999PASP..111...84V}{Vanture} \& {Wallerstein} (1999). They derived their O
abundance with measured EWs for the O triplet. The abundances were corrected for NLTE effects with
the prescription of \nocite{1979A&A....71..178E}{Eriksson} \& {Toft} (1979). The difference in \Teff
between ours and the determination of Vanture et al. might account for the major part of the
difference. Raising our \Teff by 400\,K would lead to O = 8.71. 

\subsection{Outliers\label{sect:outliers}} 
\begin{figure}
\centering
\includegraphics[width=\linewidth]{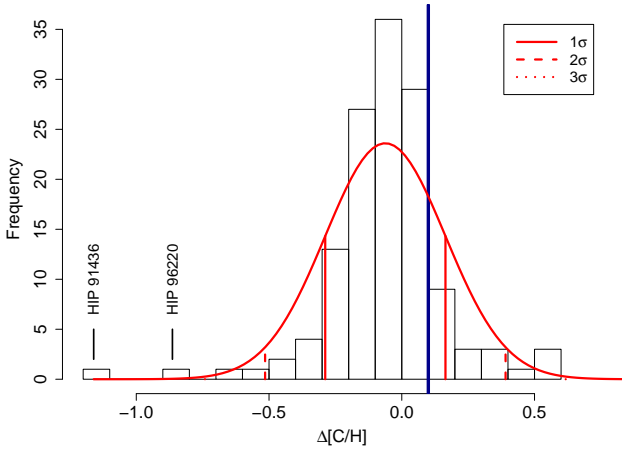}
\caption{Histogram of differences in [C/H] from assumed offset of -0.1\,dex from
\protect\nocite{2006AJ....131.3069L}{Luck} \& {Heiter} (2006). The vertical line indicates the trend
of Luck \& Heiter.  \label{fig:f8}} 
\end{figure}

To get a quantitative estimate of the divergence in C abundances of the program stars in Figure
\ref{fig:f5}, the difference between [C/H] of each star and the assumed offset of -0.1\,dex from the
trend of \nocite{2006AJ....131.3069L}{Luck} \& {Heiter} (2006) was determined (see Section
\ref{sect:carbon}), i.e. how much the [C/H] abundances diverge from the dashed line in Figure
\ref{fig:f5}. To account for the uncertainty in the abundance determination these differences
$\Delta$[C/H] were multiplied with a random value of the order of the mean error in [C/H]. This
increases the scatter in the distribution of $\Delta$[C/H] and helps to detect if the distribution
of outliers is in agreement with the observational scatter. A histogram of $\Delta$[C/H] can be seen
in Figure \ref{fig:f8}. It shows that the distribution is centered on zero, as expected, and falls
off steeply.  The [C/H] abundances of most stars in the sample are therefore in accordance with the
observational scatter about the expected trend with [Fe/H].

\begin{figure}
\centering
\includegraphics[width=\linewidth]{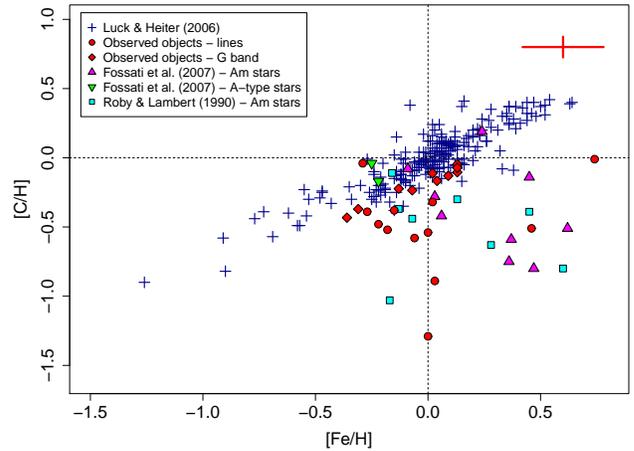}
\caption{C abundances of Am/Ap stars in the sample compared with literature values of a different
set of Am/Ap stars and normal stars.\label{fig:f9}} 
\end{figure}

However, two objects show a $\Delta \text{[C/H]}$ larger than $3\,\sigma$,
HIP\,91436 and HIP\,96220. Both stars belong to the class of the Am and Ap
stars.  Abundance peculiarities for Am and Ap stars are well known and usually
appear in the form of enhanced abundances of heavy metals and lower abundances
of elements like Ca or Sc. A search through the most extensive catalogue of Am,
Ap, and HgMn stars by \nocite{2009A&A...498..961R}{Renson} \& {Manfroid} (2009)
revealed that 28 stars in our sample have previously been identified as Am or
Ap stars including the majority of stars with low C abundances. In Figure
\ref{fig:f9} only the Am or Ap stars in the sample are plotted and compared to
C determinations for a sample of Am and Ap stars from
\nocite{2007A&A...476..911F}{Fossati} {et~al.} (2007) and
\nocite{1990ApJS...73...67R}{Roby} \& {Lambert} (1990).  For the objects in
Roby \& Lambert no metallicities were provided by the authors and the values
for [Fe/H] in Figure \ref{fig:f9} were taken from other literature sources. The
underabundances in C for their objects are of comparable magnitude as the ones
found for the Am/Ap stars among the observed objects. In addition to the C
deficiency, the literature abundances generally show a deficiency in O that is
not apparent in our Am/Ap sample (compare Figure \ref{fig:f7}). The comparison
with Figure \ref{fig:f5} and \ref{fig:f8} also shows that the Am/Ap stars
account for all stars with low [C/H] and $\Delta$[C/H]$> 2\,\sigma$.

\section{Discussion}
\begin{figure}
\centering
\includegraphics[width=\linewidth]{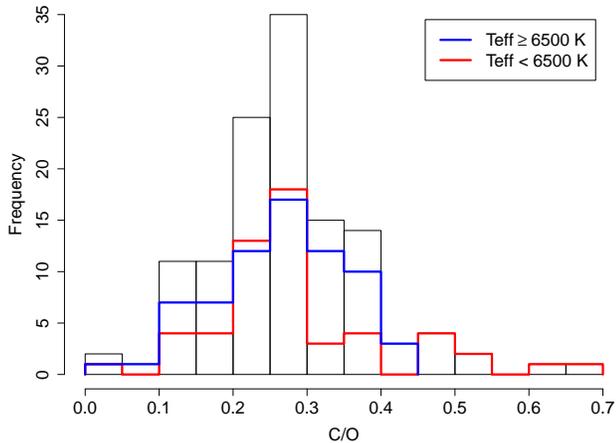}
\caption{C/O abundances for our program stars. \label{fig:f10}}
\end{figure}

\begin{figure}
\centering
\includegraphics[width=\linewidth]{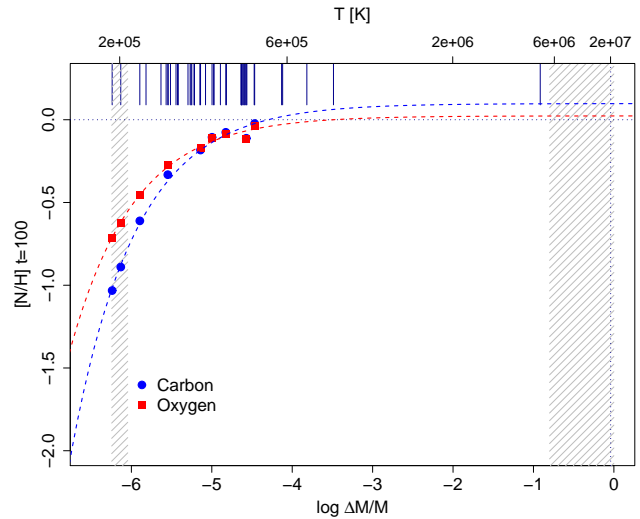}
\caption{Surface abundance for C and O for a 2.5\,\Msun model vs. depths of turbulent
mixing. Dashed regions mark convection zones below the surface and in the core. Vertical lines at
the top of the plot mark mixing depths of all turbulence models presented in
\protect\nocite{2000ApJ...529..338R}{Richer} {et~al.} (2000). \label{fig:f11}} 
\end{figure}

Our analysis of a large sample of stars between 2 -- 5\,\Msun in the Hertzsprung gap shows that the
abundances of C and O are generally consistent with those of normal mass dwarf stars and that no
significant alteration of these elements can be found before the atmosphere is homogenized by the
first dredge-up and abundance changes for light nuclides are introduced as material exposed to mild
H-burning is mixed into the deep convective envelope. Before the dredge-up the abundances of C and O
are stable during the evolution through the Hertzsprung gap. This is evident by the constant C/O
ratio during this period that shows no dependency on \Teff \, as can be seen in Figure \ref{fig:f10}.
The fact that no difference in the C/O
ratio can be found between stars located on the two sides of the Hertzsprung gap indicates that any
rotationally induced mixing that might be present in stars with high rotational velocities at the
blue side of the Hertzsprung gap is not strong enough to have an effect on the relative abundances
of C and O in these stars.  In fact, the only abundance anomalies in the sample are apparent for the
slow rotating Am/Ap stars. As noted above, the Am and Ap stars generally exhibit a C deficiency.
This abundance anomaly and the other numerous anomalies found in other Am/Ap stars are widely
attributed to diffusive separation in the outer envelope of the star resulting from competition
between gravitational settling and radiative levitation \nocite{1970ApJ...160..641M}({Michaud}
1970).  Successful diffusive separation requires slow rotation rates to suppress
rotationally-induced mixing and this explains why the majority of slow rotators among A-type main
sequence stars are chemically peculiar \nocite{2009AJ....138...28A}({Abt} 2009).  In terms of the
light elements, the C, N, and O abundances found by \nocite{2007A&A...476..911F}{Fossati} {et~al.}
(2007) can be reproduced well by the theoretical realization of diffusive separation by
\nocite{2000ApJ...529..338R}{Richer}, {Michaud}, \& {Turcotte} (2000).  Since the Am and Ap
abundance anomalies result from a redistribution of an element within the envelope including the
atmosphere, the redistribution is canceled by the deep convective envelope that constitutes the
first dredge-up. In the models of \nocite{2000ApJ...529..338R}{Richer} {et~al.} (2000) the strength
of the abundance anomaly in the atmosphere is controlled by the depth of the zone mixed by
turbulence. For the observed abundance anomalies of the Am/Ap stars the central regions of the star
are not mixed significantly. That suggests that the mechanism used to explain the Am/Ap star
abundances can not be responsible for the C deficiency found in the peculiar wk Gb stars that
require that CN-cycled material was processed at temperatures of 20 million degrees or hotter in the
interior of the star \nocite{2013ApJ...765..155A}({Adamczak} \& {Lambert} 2013, see Figure
\ref{fig:f11}). For these objects rotation rates even higher than observed for the stars in the
present sample might be required to induce mixing that extends into interior regions where some
H-burning occurs and to connect these regions with the surface. Some rapidly-rotating stars might
then be C-poor and N-rich. A search for such stars will be challenging but realizable
\nocite{1986PASP...98..927L}({Lambert}, {McKinley}, \&  {Roby} 1986).

\begin{figure}
\centering
\includegraphics[width=\linewidth]{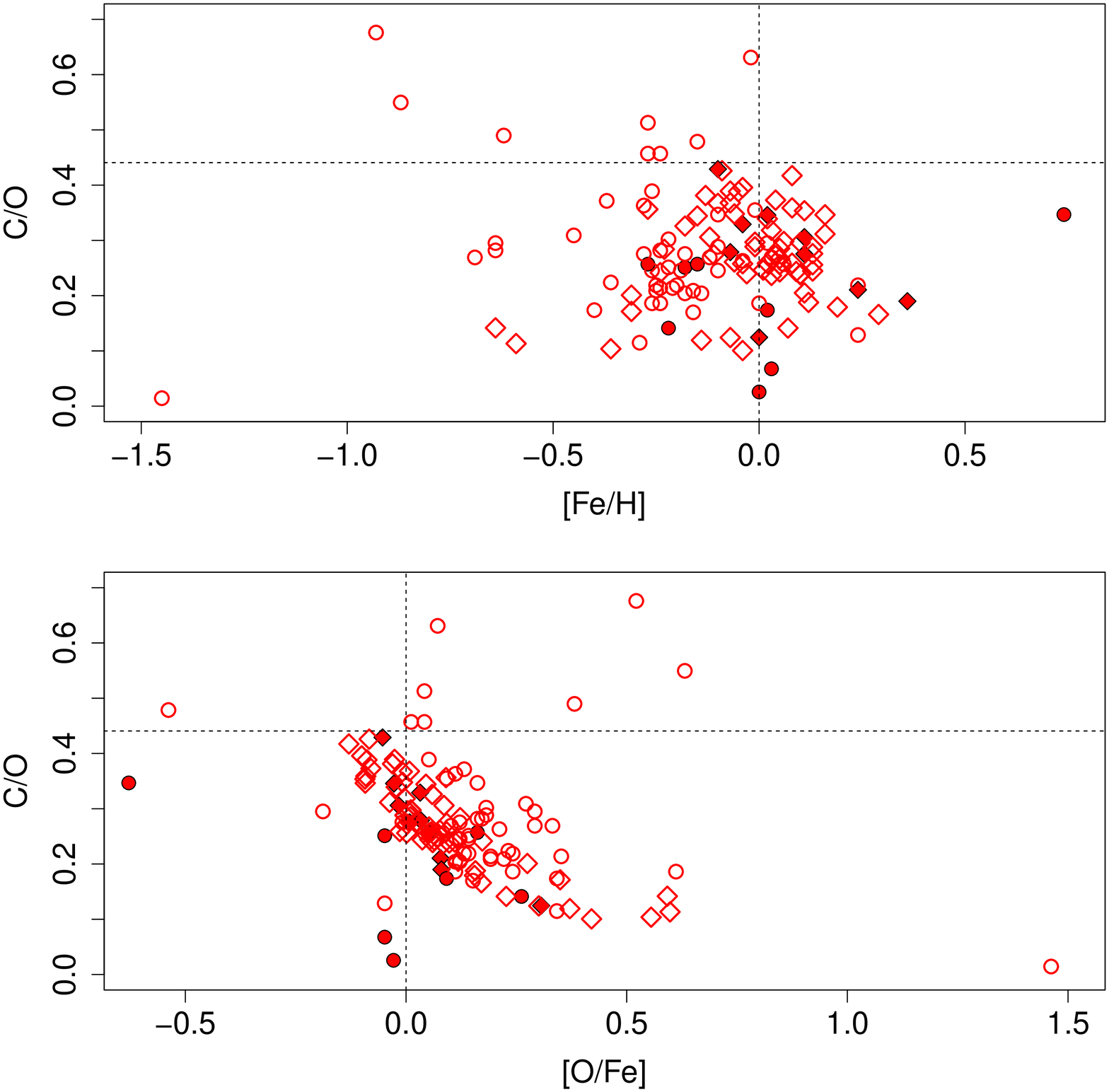}
\caption{C/O ratio for our program stars vs. [Fe/H] (top panel) and [O/Fe] (bottom panel). The
symbol shapes are the same as in Figures \ref{fig:f5} and \ref{fig:f7}, filled symbols mark the
Am/Ap stars in the sample. \label{fig:f12}} 
\end{figure}
Apart from the peculiar stars in the sample, the abundances of the stars in the Hertzsprung gap
follow the systematics found for normal mass stars. Furthermore, the abundances of all stars are in
accordance with the Galactic chemical evolution. There is a trend between the C/O ratio and the
[O/Fe] abundance of the stars in our sample and the stars with low metallicities tend to have higher
[O/Fe] (see bottom panel of Figure \ref{fig:f12}). \nocite{1999A&A...342..426G}{Gustafsson} {et~al.}
(1999) found a trend between the C/O ratio and [Fe/H] for late F and early G type dwarfs in the
Galactic Disk that they attributed to the metallicity dependence of the yields of C production in
massive stars and the difference in time scale between the C production in less massive stars and
the O production in massive stars. We can not confirm such a trend for the stars in our sample.
There is evidence that the stars in the Hertzsprung gap are in agreement with a rather constant
relation between C/O and [Fe/H]. This includes the Am/Ap stars that can not be clearly distinguished
from the rest of the sample (cf. top panel in Figure \ref{fig:f12}). The local dwarfs from
\nocite{2006AJ....131.3069L}{Luck} \& {Heiter} (2006) show a similar behavior and scatter with no
certain trend. Compared to their objects, however, the C/O ratio for most of the program stars is
subsolar. This is a reflection of the slightly lower C abundances obtained by our abundance
analysis (see Section \ref{sect:carbon}). 

\section{Concluding Remarks}
Our C and O abundance analysis of a large sample of stars in the Hertzsprung gap allows to draw a
picture of the characteristics of stars in the mass range 2 -- 5\,\Msun in their evolution from the
main sequence to the giant branch. Apart from the chemical peculiarities of the Am/Ap stars that are
formed in their main sequence phase and are erased by the first dredge-up as the stars ascend the
giant branch, it appears that no other abundance anomalies are surfacing during this evolutionary
state and that the abundances of basic elements like C and O are not noticeably altered during the
transition of the gap before the dredge-up. 

The abundances determined for the program stars are in agreement with the those of local lower mass
dwarf stars and follow the trends that are expected from Galactic chemical evolution. No clear
indication of severe rotationally-induced mixing in the fast rotating stars of the sample could be
found, however, we are limited here by the methods applied and an extension of the abundance
analysis towards higher rotation rates might be worthwhile.

\acknowledgments

We thank the referee for constructive comments and suggestions.
This research has been supported in part by the Robert A. Welch Foundation of Houston, Texas through
grant F-634.


\begin{thebibliography}{}

\bibitem[{Abt} 2009]{2009AJ....138...28A}
{Abt}, H.~A. 2009, \aj, 138, 28

\bibitem[{Abt} \& {Morrell} 1995]{1995ApJS...99..135A}
{Abt}, H.~A. \& {Morrell}, N.~I. 1995, \apjs, 99, 135

\bibitem[{Adamczak} \& {Lambert} 2013]{2013ApJ...765..155A}
{Adamczak}, J. \& {Lambert}, D.~L. 2013, \apj, 765, 155

\bibitem[{Adelman}, {Cay}, {Tektunali}, {Gulliver}, \&  {Teker} 2008]{2008AN....329....4A}
{Adelman}, S.~J., {Cay}, I.~H., {Tektunali}, H.~G., {Gulliver}, A.~F., \&  {Teker}, A. 2008, Astronomische Nachrichten, 329, 4

\bibitem[{Allende Prieto}, {Lambert}, \&  {Asplund} 2001]{2001ApJ...556L..63A}
{Allende Prieto}, C., {Lambert}, D.~L., \& {Asplund}, M. 2001, \apjl, 556, L63

\bibitem[{Alonso}, {Arribas}, \&  {Mart{\'{\i}}nez-Roger} 1999]{1999A&AS..140..261A}
{Alonso}, A., {Arribas}, S., \& {Mart{\'{\i}}nez-Roger}, C. 1999, \aaps, 140,  261

\bibitem[{Andrievsky}, {Kovtyukh}, {Luck},  {L{\'e}pine}, {Bersier}, {Maciel}, {Barbuy}, {Klochkova}, {Panchuk}, \&  {Karpischek} 2002]{2002A&A...381...32A}
{Andrievsky}, S.~M., {Kovtyukh}, V.~V., {Luck}, R.~E., {L{\'e}pine}, J.~R.~D.,  {Bersier}, D., {Maciel}, W.~J., {Barbuy}, B., {Klochkova}, V.~G., {et al.} 2002, \aap, 381, 32

\bibitem[{Asplund}, {Grevesse}, {Sauval}, {Allende  Prieto}, \& {Blomme} 2005]{2005A&A...431..693A}
{Asplund}, M., {Grevesse}, N., {Sauval}, A.~J., {Allende Prieto}, C., \&  {Blomme}, R. 2005, \aap, 431, 693

\bibitem[{Asplund}, {Grevesse}, {Sauval}, {Allende  Prieto}, \& {Kiselman} 2004]{2004A&A...417..751A}
{Asplund}, M., {Grevesse}, N., {Sauval}, A.~J., {Allende Prieto}, C., \&  {Kiselman}, D. 2004, \aap, 417, 751

\bibitem[{Asplund}, {Grevesse}, {Sauval}, \&  {Scott} 2009]{2009ARA&A..47..481A}
{Asplund}, M., {Grevesse}, N., {Sauval}, A.~J., \& {Scott}, P. 2009, \araa, 47,  481

\bibitem[{Balega}, {Leushin}, {Kuznetsov}, \&  {Tamazian} 2008]{2008ARep...52..226B}
{Balega}, Y.~Y., {Leushin}, V.~V., {Kuznetsov}, M.~K., \& {Tamazian}, V. 2008,  Astronomy Reports, 52, 226

\bibitem[{Barklem}, {Stempels}, {Allende Prieto},  {Kochukhov}, {Piskunov}, \& {O'Mara} 2002]{2002A&A...385..951B}
{Barklem}, P.~S., {Stempels}, H.~C., {Allende Prieto}, C., {Kochukhov}, O.~P.,  {Piskunov}, N., \& {O'Mara}, B.~J. 2002, \aap, 385, 951

\bibitem[{Bernacca} \& {Perinotto} 1970]{1970CoAsi.239....1B}
{Bernacca}, P.~L. \& {Perinotto}, M. 1970, Contributions dell'Osservatorio  Astrofisica dell'Universita di Padova in Asiago, 239, 1

\bibitem[{Bertelli}, {Girardi}, {Marigo}, \&  {Nasi} 2008]{2008A&A...484..815B}
{Bertelli}, G., {Girardi}, L., {Marigo}, P., \& {Nasi}, E. 2008, \aap, 484, 815

\bibitem[{Bertelli}, {Nasi}, {Girardi}, \&  {Marigo} 2009]{2009A&A...508..355B}
{Bertelli}, G., {Nasi}, E., {Girardi}, L., \& {Marigo}, P. 2009, \aap, 508, 355

\bibitem[{Bidelman} 1951]{1951ApJ...113..304B}
{Bidelman}, W.~P. 1951, \apj, 113, 304

\bibitem[{Cameron} \& {Fowler} 1971]{1971ApJ...164..111C}
{Cameron}, A.~G.~W. \& {Fowler}, W.~A. 1971, \apj, 164, 111

\bibitem[{Carquillat} \& {Prieur} 2007]{2007MNRAS.380.1064C}
{Carquillat}, J.-M. \& {Prieur}, J.-L. 2007, \mnras, 380, 1064

\bibitem[{Castelli} \& {Kurucz} 2004]{2004astro.ph..5087C}
{Castelli}, F. \& {Kurucz}, R.~L. 2004, Modeling of Stellar Atmospheres (IAU Symp. No. 210), ed. N.
Piskunov, W. Weiss, \& D. Gray, 2003, poster A20 (arXiv:astro-ph/0405087)

\bibitem[{Cottrell} \& {Norris} 1978]{1978ApJ...221..893C}
{Cottrell}, P.~L. \& {Norris}, J. 1978, \apj, 221, 893

\bibitem[{de Laverny}, {do Nascimento},  {L{\`e}bre}, \& {De Medeiros} 2003]{2003A&A...410..937D}
{de Laverny}, P., {do Nascimento}, Jr., J.~D., {L{\`e}bre}, A., \& {De  Medeiros}, J.~R. 2003, \aap, 410, 937

\bibitem[{Demarque}, {Woo}, {Kim}, \&  {Yi} 2004]{2004ApJS..155..667D}
{Demarque}, P., {Woo}, J.-H., {Kim}, Y.-C., \& {Yi}, S.~K. 2004, \apjs, 155,  667

\bibitem[{Ecuvillon}, {Israelian}, {Santos},  {Mayor}, {Villar}, \& {Bihain} 2004]{2004A&A...426..619E}
{Ecuvillon}, A., {Israelian}, G., {Santos}, N.~C., {Mayor}, M., {Villar}, V.,  \& {Bihain}, G. 2004, \aap, 426, 619

\bibitem[{Ecuvillon}, {Israelian}, {Santos},  {Shchukina}, {Mayor}, \& {Rebolo} 2006]{2006A&A...445..633E}
{Ecuvillon}, A., {Israelian}, G., {Santos}, N.~C., {Shchukina}, N.~G., {Mayor},  M., \& {Rebolo}, R. 2006, \aap, 445, 633

\bibitem[{Eriksson} \& {Toft} 1979]{1979A&A....71..178E}
{Eriksson}, K. \& {Toft}, S.~C. 1979, \aap, 71, 178

\bibitem[{ESA} 1997]{1997yCat.1239....0E}
{ESA}. 1997, VizieR Online Data Catalog, 1239, 0

\bibitem[{Fossati}, {Bagnulo}, {Monier}, {Khan},  {Kochukhov}, {Landstreet}, {Wade}, \& {Weiss} 2007]{2007A&A...476..911F}
{Fossati}, L., {Bagnulo}, S., {Monier}, R., {Khan}, S.~A., {Kochukhov}, O.,  {Landstreet}, J., {Wade}, G., \& {Weiss}, W. 2007, \aap, 476, 911

\bibitem[{Galeev}, {Ivanova}, {Shimansky}, \&  {Bikmaev} 2012]{2012ARep...56..850G}
{Galeev}, A.~I., {Ivanova}, D.~V., {Shimansky}, V.~V., \& {Bikmaev}, I.~F.  2012, Astronomy Reports, 56, 850

\bibitem[{Gustafsson}, {Karlsson}, {Olsson}, {Edvardsson}, \& {Ryde} 1999]{1999A&A...342..426G}
{Gustafsson}, B., {Karlsson}, T., {Olsson}, E., {Edvardsson}, B., \& {Ryde}, N. 1999, \aap, 342, 426

\bibitem[{Hakkila}, {Myers}, {Stidham}, \&  {Hartmann} 1997]{1997AJ....114.2043H}
{Hakkila}, J., {Myers}, J.~M., {Stidham}, B.~J., \& {Hartmann}, D.~H. 1997,  \aj, 114, 2043

\bibitem[{Hiltgen} 1996]{1996PhDT........75H}
{Hiltgen}, D.~D. 1996, PhD thesis, PhD thesis, Univ.~Texas Austin.~ (1996)

\bibitem[{Johansson}, {Litz{\'e}n}, {Lundberg}, \&  {Zhang} 2003]{2003ApJ...584L.107J}
{Johansson}, S., {Litz{\'e}n}, U., {Lundberg}, H., \& {Zhang}, Z. 2003, \apjl,  584, L107

\bibitem[{Kolenberg}, {Fossati}, {Shulyak},  {Pikall}, {Barnes}, {Kochukhov}, \& {Tsymbal} 2010]{2010A&A...519A..64K}
{Kolenberg}, K., {Fossati}, L., {Shulyak}, D., {Pikall}, H., {Barnes}, T.~G.,  {Kochukhov}, O., \& {Tsymbal}, V. 2010, \aap, 519, A64

\bibitem[{Koleva}, {Prugniel}, {Bouchard}, \&  {Wu} 2009]{2009A&A...501.1269K}
{Koleva}, M., {Prugniel}, P., {Bouchard}, A., \& {Wu}, Y. 2009, \aap, 501, 1269

\bibitem[{Kraft} 1967]{1967ApJ...150..551K}
{Kraft}, R.~P. 1967, \apj, 150, 551

\bibitem[{Kurucz}, {Furenlid}, {Brault}, \&  {Testerman} 1984]{1984sfat.book.....K}
{Kurucz}, R.~L., {Furenlid}, I., {Brault}, J., \& {Testerman}, L. 1984, {Solar  flux atlas from 296 to 1300 nm}

\bibitem[{Lambert}, {McKinley}, \&  {Roby} 1986]{1986PASP...98..927L}
{Lambert}, D.~L., {McKinley}, L.~K., \& {Roby}, S.~W. 1986, \pasp, 98, 927

\bibitem[{Lambert} \& {Sawyer} 1984]{1984ApJ...283..192L}
{Lambert}, D.~L. \& {Sawyer}, S.~R. 1984, \apj, 283, 192

\bibitem[{Lind}, {Bergemann}, \&  {Asplund} 2012]{2012MNRAS.427...50L}
{Lind}, K., {Bergemann}, M., \& {Asplund}, M. 2012, \mnras, 427, 50

\bibitem[{Luck}, {Andrievsky}, {Fokin}, \&  {Kovtyukh} 2008]{2008AJ....136...98L}
{Luck}, R.~E., {Andrievsky}, S.~M., {Fokin}, A., \& {Kovtyukh}, V.~V. 2008,  \aj, 136, 98

\bibitem[{Luck} \& {Heiter} 2006]{2006AJ....131.3069L}
{Luck}, R.~E. \& {Heiter}, U. 2006, \aj, 131, 3069

\bibitem[{Luck} \& {Heiter} 2007]{2007AJ....133.2464L}
---. 2007, \aj, 133, 2464

\bibitem[{Luck} \& {Wepfer} 1995]{1995AJ....110.2425L}
{Luck}, R.~E. \& {Wepfer}, G.~G. 1995, \aj, 110, 2425

\bibitem[{Massarotti}, {Latham}, {Stefanik}, \&  {Fogel} 2008]{2008AJ....135..209M}
{Massarotti}, A., {Latham}, D.~W., {Stefanik}, R.~P., \& {Fogel}, J. 2008, \aj,  135, 209

\bibitem[{Mermilliod}, {Mermilliod}, \&  {Hauck} 1997]{1997A&AS..124..349M}
{Mermilliod}, J.-C., {Mermilliod}, M., \& {Hauck}, B. 1997, \aaps, 124, 349

\bibitem[{Michaud} 1970]{1970ApJ...160..641M}
{Michaud}, G. 1970, \apj, 160, 641

\bibitem[{Michaud} 1977]{1977Natur.266..433M}
---. 1977, \nat, 266, 433

\bibitem[{Palacios}, {Parthasarathy}, {Bharat Kumar},  \& {Jasniewicz} 2012]{2012A&A...538A..68P}
{Palacios}, A., {Parthasarathy}, M., {Bharat Kumar}, Y., \& {Jasniewicz}, G.  2012, \aap, 538, A68

\bibitem[{Prugniel} \& {Soubiran} 2001]{2001yCat.3218....0P}
{Prugniel}, P. \& {Soubiran}, C. 2001, VizieR Online Data Catalog, 3218, 0

\bibitem[{Prugniel}, {Vauglin}, {I.}, \&  {Koleva} 2011]{2011yCat..35319165P}
{Prugniel}, P., {Vauglin}, {I.}, \& {Koleva}, M. 2011, VizieR Online Data  Catalog, 353, 19165

\bibitem[{Przybilla}, {Butler}, \&  {Kudritzki} 2001]{2001A&A...379..936P}
{Przybilla}, N., {Butler}, K., \& {Kudritzki}, R.~P. 2001, \aap, 379, 936

\bibitem[{Rachkovskaya} 2003]{2003ARep...47..865R}
{Rachkovskaya}, T.~M. 2003, Astronomy Reports, 47, 865

\bibitem[{Ram{\'{\i}}rez} \& {Allende Prieto} 2011]{2011ApJ...743..135R}
{Ram{\'{\i}}rez}, I. \& {Allende Prieto}, C. 2011, \apj, 743, 135

\bibitem[{Ram{\'{\i}}rez}, {Allende Prieto}, \&  {Lambert} 2007]{2007A&A...465..271R}
{Ram{\'{\i}}rez}, I., {Allende Prieto}, C., \& {Lambert}, D.~L. 2007, \aap,  465, 271

\bibitem[{Ram{\'{\i}}rez} \& {Mel{\'e}ndez} 2005]{2005ApJ...626..465R}
{Ram{\'{\i}}rez}, I. \& {Mel{\'e}ndez}, J. 2005, \apj, 626, 465

\bibitem[{Renson} \& {Manfroid} 2009]{2009A&A...498..961R}
{Renson}, P. \& {Manfroid}, J. 2009, \aap, 498, 961

\bibitem[{Richard}, {Michaud}, \&  {Richer} 2001]{2001ApJ...558..377R}
{Richard}, O., {Michaud}, G., \& {Richer}, J. 2001, \apj, 558, 377

\bibitem[{Richard}, {Michaud}, {Richer}, {Turcotte},  {Turck-Chi{\`e}ze}, \& {VandenBerg} 2002]{2002ApJ...568..979R}
{Richard}, O., {Michaud}, G., {Richer}, J., {Turcotte}, S., {Turck-Chi{\`e}ze},  S., \& {VandenBerg}, D.~A. 2002, \apj, 568, 979

\bibitem[{Richer}, {Michaud}, \&  {Turcotte} 2000]{2000ApJ...529..338R}
{Richer}, J., {Michaud}, G., \& {Turcotte}, S. 2000, \apj, 529, 338

\bibitem[{Roby} \& {Lambert} 1990]{1990ApJS...73...67R}
{Roby}, S.~W. \& {Lambert}, D.~L. 1990, \apjs, 73, 67

\bibitem[{Royer}, {Grenier}, {Baylac}, {G{\'o}mez}, \&  {Zorec} 2002]{2002A&A...393..897R}
{Royer}, F., {Grenier}, S., {Baylac}, M.-O., {G{\'o}mez}, A.~E., \& {Zorec}, J.  2002, \aap, 393, 897

\bibitem[{Russell} 1995]{1995ApJ...451..747R}
{Russell}, S.~C. 1995, \apj, 451, 747

\bibitem[{Saffe} 2011]{2011RMxAA..47....3S}
{Saffe}, C. 2011, rmxaa, 47, 3

\bibitem[{Schr{\"o}der}, {Reiners}, \&  {Schmitt} 2009]{2009A&A...493.1099S}
{Schr{\"o}der}, C., {Reiners}, A., \& {Schmitt}, J.~H.~M.~M. 2009, \aap, 493,  1099

\bibitem[{Sneden}, {Lambert}, {Tomkin}, \&  {Peterson} 1978]{1978ApJ...222..585S}
{Sneden}, C., {Lambert}, D.~L., {Tomkin}, J., \& {Peterson}, R.~C. 1978, \apj,  222, 585

\bibitem[{Sneden} 1973]{1973PhDT.......180S}
{Sneden}, C.~A. 1973, PhD thesis, The University of Texas at Austin.

\bibitem[{Takeda}, {Honda}, {Aoki}, {Takada-Hidai},  {Zhao}, {Chen}, \& {Shi} 2006]{2006PASJ...58..389T}
{Takeda}, Y., {Honda}, S., {Aoki}, W., {Takada-Hidai}, M., {Zhao}, G., {Chen},  Y.-Q., \& {Shi}, J.-R. 2006, \pasj, 58, 389

\bibitem[{Takeda}, {Kang}, {Han}, {Lee}, \&  {Kim} 2013]{2013MNRAS.432..769T}
{Takeda}, Y., {Kang}, D.-I., {Han}, I., {Lee}, B.-C., \& {Kim}, K.-M. 2013,  \mnras, 432, 769

\bibitem[{Takeda}, {Sato}, \&  {Murata} 2008]{2008PASJ...60..781T}
{Takeda}, Y., {Sato}, B., \& {Murata}, D. 2008, \pasj, 60, 781

\bibitem[{Takeda}, {Takada-Hidai}, {Jugaku}, {Sakaue},  \& {Sadakane} 1999]{1999PASJ...51..961T}
{Takeda}, Y., {Takada-Hidai}, M., {Jugaku}, J., {Sakaue}, A., \& {Sadakane}, K.  1999, \pasj, 51, 961

\bibitem[{Tomkin}, {Sneden}, \&  {Cottrell} 1984]{1984PASP...96..609T}
{Tomkin}, J., {Sneden}, C., \& {Cottrell}, P.~L. 1984, \pasp, 96, 609

\bibitem[{Tull}, {MacQueen}, {Sneden}, \&  {Lambert} 1995]{1995PASP..107..251T}
{Tull}, R.~G., {MacQueen}, P.~J., {Sneden}, C., \& {Lambert}, D.~L. 1995,  \pasp, 107, 251

\bibitem[{van Leeuwen} 2007]{2007A&A...474..653V}
{van Leeuwen}, F. 2007, \aap, 474, 653

\bibitem[{Vanture} \& {Wallerstein} 1999]{1999PASP..111...84V}
{Vanture}, A.~D. \& {Wallerstein}, G. 1999, \pasp, 111, 84

\bibitem[{Wallerstein}, {B\"ohm-Vitense},  {Vanture}, \& {Gonzalez} 1994]{1994AJ....107.2211W}
{Wallerstein}, G., {B\"ohm-Vitense}, E., {Vanture}, A.~D., \& {Gonzalez}, G.  1994, \aj, 107, 2211

\bibitem[{Wu}, {Singh}, {Prugniel}, {Gupta}, \&  {Koleva} 2011]{2011A&A...525A..71W}
{Wu}, Y., {Singh}, H.~P., {Prugniel}, P., {Gupta}, R., \& {Koleva}, M. 2011,  \aap, 525, A71

\end{thebibliography}
\end{document}